UNIVERSITY OF LATVIA

VALDIS VITOLINS

# BUSINESS PROCESS MODELING USING A METAMODELING APPROACH

Summary of Doctoral Thesis

Advisor:
Professor, Dr. habil. sc. comp.
**AUDRIS KALNIŅŠ**

The thesis was elaborated with support of ESF
(European Social Fund)

Advisor:

*Professor, Dr. habil. sc. comp. Audris Kalnins*
*University of Latvia*

Referees:

*Professor, Dr. habil. sc. comp. Janis Barzdins*
*University of Latvia*

*Assoc. Professor, Dr. sc. ing. Marite Kirikova*
*Riga Technical University*

*Professor, Ph. D. Olegas Vasilecas*
*Vilnius Gediminas Technical University*

The defense of the thesis will take place in an open session of the Council for Promotion in Computer Science, the University of Latvia, in the Institute of Mathematics and Computer Science, the University of Latvia *(Room 413, Raiņa bulv. 29, Rīga), on 11. September 2007.*

The thesis (collection of works) and its summary are available at the Library of the University *(Kalpaka bulv. 4, Rīga)* .

Head of the Council Janis Barzdins

# Content

# Figures

# Relevance of the Thesis and Achieved Results

**Relevance of the thesis.** The thesis discusses topics related to the development of business process management systems. Business process management systems have evolved on the basis of workflow management systems through incremental inclusion of standard information system functions, for example, resource and client management.

The application of model driven development is required to deal with the complexity of business management systems and to increase development efficiency. In contrast to conventional information systems, the behavior of business management systems is strongly affected by the business models that they execute. Thus, business process models also can be used for designing and developing business management systems using sequentially applied model transformations that adapt models to a specific execution platform.

As business management system functionality expands and business models are used for systems development, new requirements emerged for both business models and process modeling languages. Currently, there is no agreement on a single best business process modeling language; therefore, different aspects of business management systems in different development phases are described using different modeling languages.

For example, the *de facto* standard in software development, the Unified Modeling Language (UML), is not detailed enough to use in workflow modeling. The Business Process Modeling Notation (BPMN) is not explicit enough to use for software development. The execution semantics of existing languages is not detailed enough for process execution in distributed business management systems. This makes it difficult to use models in the development of business process management systems, and currently model driven development can be applied only to specific business management branches and only in specific development stages.

Therefore, it is necessary to develop an approach that supports the description of different business management aspects in different modeling languages, and which supports the transition from one language to another, both to apply a business model to a specific execution platform, and to move to an alternative modeling language. This approach could support a unified tool platform that would cover all required business process management system aspects, and information gathered in different development stages could be effectively reused.

This thesis proposes that business process modeling problems can be solved using a metamodeling approach. Metamodeling allows analysis of different modeling aspects in a unified and comprehensive way, while retaining the exact semantics of concepts. In this thesis, a methodology is developed that allows analyzing different modeling languages and comparing their concepts in detail. The thesis also describes model execution and the development of run-time measurement techniques using a metamodeling approach. Based on the developed methodology, a framework with editors for different modeling languages and for model transformations that transform concrete business models for a specific execution platform is developed.

**The main results of the research.**

- ❑ A notation independent business process metamodel is developed, which shows business process concepts and their relationships. This metamodel is used as a canonical form of business process modeling languages in all further research.

- A new approach is developed for mapping business concepts from one common domain to different modeling languages using similar concepts ("semantically similar" languages). This approach is used as the basis for building model transformations.
- Exact execution semantics for the UML activity diagram (AD) is developed using a virtual machine. A methodology based on metamodels is developed. It allows defining model parameters in design-time and measuring them in model run-time. This virtual machine can be used as the basis for developing a process simulation or business process management system.
- Functionally equivalent metamodels of the most popular business modeling languages, a profile of the UML AD subset and a BPMN subset, are developed. On the basis of these metamodels, editors for these languages are developed. Model transformations are developed in the MOdeling LAnguage (MOLA) to perform transformations from AD to BPMN. Such transformations are one step in the model driven development of business management systems. By using a similar approach to transform models further into the Business Process Execution Language (BPEL), the models can be executed in real business process management systems.

## General Description of the Thesis

Research on the thesis **"Business Process Modeling Using a Metamodeling Approach"** was done from 2002 to 2007 in the Department of Physics and Mathematics of the University of Latvia and in the Institute of Mathematics and Computer Science (IMCS) under the direction of Professor Audris Kalnins. This research is a continuation of the business modeling traditions established in the IMCS since 1986.

The main results of the research are published in four papers [1-4] and are presented at four international conferences. **The thesis is organized as the set of these four papers**, summarizing the author's research results in different aspects of business modeling.

**The subject of the research:** business models, comparison of different models and modeling techniques, and analysis of model execution using the metamodeling approach.

**The goals of the research:** to develop a comprehensive and exact approach to using different business process modeling aspects, process execution, measurement and transformations, using the metamodeling approach.

**Research stages:** research was done in several stages, starting with a general analysis of different business modeling aspects, and continuing with exact semantics of modeling languages and their execution semantics. Using exact semantics of concepts, transformations were developed to translate business models to a language executable in business process management systems. In the subsequent sections of this summary, the main problem statements are briefly described. These are described in detail in the referenced publications.

- In the first chapter **"Models metamodels and metamodeling"** main concepts of metamodeling are described. This chapter explains approaches and standards – namely UML modeling language and MOF, which is later used everywhere in the paper. In this chapter relation to MOF metalayers particularly to business modeling is described and equivalence/similarity of (meta)models is described in general terms.
- In the second chapter, "**Modeling Business,**" business process concepts and their relationships are analyzed using a metamodeling approach. This chapter describes existing business modeling languages and frameworks. The research analyzes business process execution and the business process environment, and presents information required to understand business processes and quality measures. Furthermore, a new comprehensive business process metamodel is developed. Using the developed business process metamodel, a new methodology is developed in which business modeling languages using similar concepts can be shown as views of a notation independent metamodel. The developed business process metamodel is used as a canonical form of business process concepts in all further investigations.
- In the third chapter, "**Business Process Measures**", existing business measurement approaches are reviewed, and a new approach to business process measurements is introduced. The new approach allows defining business measures in an integrated way together with process definition. In addition, a new methodology for process element measure definition and measure aggregation in process execution time is developed. For measure definition, a new UML Activity Diagram (AD) profile is developed, and measure algorithms are processed using an extended UML metametamodel.

- In the fourth chapter, "**Semantics of the UML 2.0 Activity Diagram for Business Modeling Using a Virtual Machine,**" a detailed description of UML 2.0 AD semantics is provided with a new approach using a virtual machine. AD elements required for business process modeling are introduced, and diagram execution in business processes is clarified. Furthermore, a new and simplified AD execution algorithm is developed, which, however, does not lose the original execution semantics. This is done using "push" and "pull" engines, which move tokens along activity execution paths. The proposed algorithm is useful for developing AD simulation and workflow execution engines.
- In the fifth chapter, "**Use of UML and Model Transformations for Workflow Process Definitions,**" a new development approach to business management systems based on model transformations is introduced. The two most popular business modeling notations, UML ADs and BPMN, are analyzed. Required aspects of workflows are briefly discussed, and, on this basis, a natural AD profile and functionally equivalent to the BPMN subset are proposed. The semantics of both languages in the context of process execution (namely, mapping to BPEL) is also analyzed. By comparing AD and BPMN metamodels, it is shown that an exact translation from AD to BPMN is not trivial; thus, model transformations are proposed as the most effective approach for model mapping and translation. Model transformations are executed in the model transformation language MOLA, using the MOLA transformation tool.
- In the **Conclusion,** research results are analyzed in the context of the latest business modeling events, and a practical approach to business process modeling is described.

**The theoretical and practical significance of the research.**

This thesis describes the theoretical principles for using the metamodels of modeling languages to describe and compare these languages. It also describes how the models written in these languages can be transformed and executed. The developed business process metamodel and mapping approach are available for exact model transformations in semantically similar languages, but model transformations make it possible to obtain model equivalents even for cases with extremely complicated relations. A metamodel-based virtual machine can be used to develop model simulation or business process management systems that execute models in different modeling languages, and the machine allows performing model measurements in run-time. The developed metamodeling approaches and ideas are used to develop a modeling tool framework and different modeling language editors. This proves the effectiveness of the metamodeling approach.

Research results are used in development of business process editors in the Generic Modeling Framework (GMF, [89]). The GMF tool is used to develop new editors for the UML AD and BPMN modeling languages. Models created in these editors can be transformed from one to another using the MOLA transformation tool.

# 1 Models metamodels and metamodeling

Model is abstract representation of real life object or phenomenon. Model can be built using different approaches. It can be formula, graph or picture. Many of real life phenomena can be modeled using discrete modeling approach – by isolating separate objects which are related each to other with some interaction. Usually such models are shown as graphical pictures where objects are shown as some kind of box, but their interaction as lines between these boxes. Although graphical representation is not mandatory, evolution of humans has proved that people are "hardwired" to resolve graphical pictures much better than reading text. Therefore discrete modeling usually uses graphical representation of the models.

Difference between simple "visualization" in pictures and modeling is that *in modeling strict rules are used* with precise meaning (semantics) of graphical elements. Standartisation and precise meaning of graphical models not only helps to understand them better, but also provide possibility to build or execute these models in software.

There are several approaches for graphical modeling, but in software modeling the most commonly used is so called metamodeling approach, which is standardized in UML standard [12]. In metamodeling each particular model is created, relying on strict model creation rules, which is also shown as a discrete model. But, because this model do not describe things of the real life but *models* (which themselves describe things of real life) this model-of-a-model is called *metamodel*. So, when models of the models are described, modeler investigates item or phenomenon in higher abstraction layer. Each particular model, which is created according to some metamodel, is concrete application or instance of the metamodel.

In general metamodels of concrete models can be created using slightly different approaches. In this paper metamodeling is done using according to approaches defined in MOF (*Meta-Object Facility*) [16] standard which is developed by OMG (*Object Management Group*), which also standardizes UML modeling language. According to MOF, there are four steps of abstraction (layers of the models), which are called from M0 to M3.

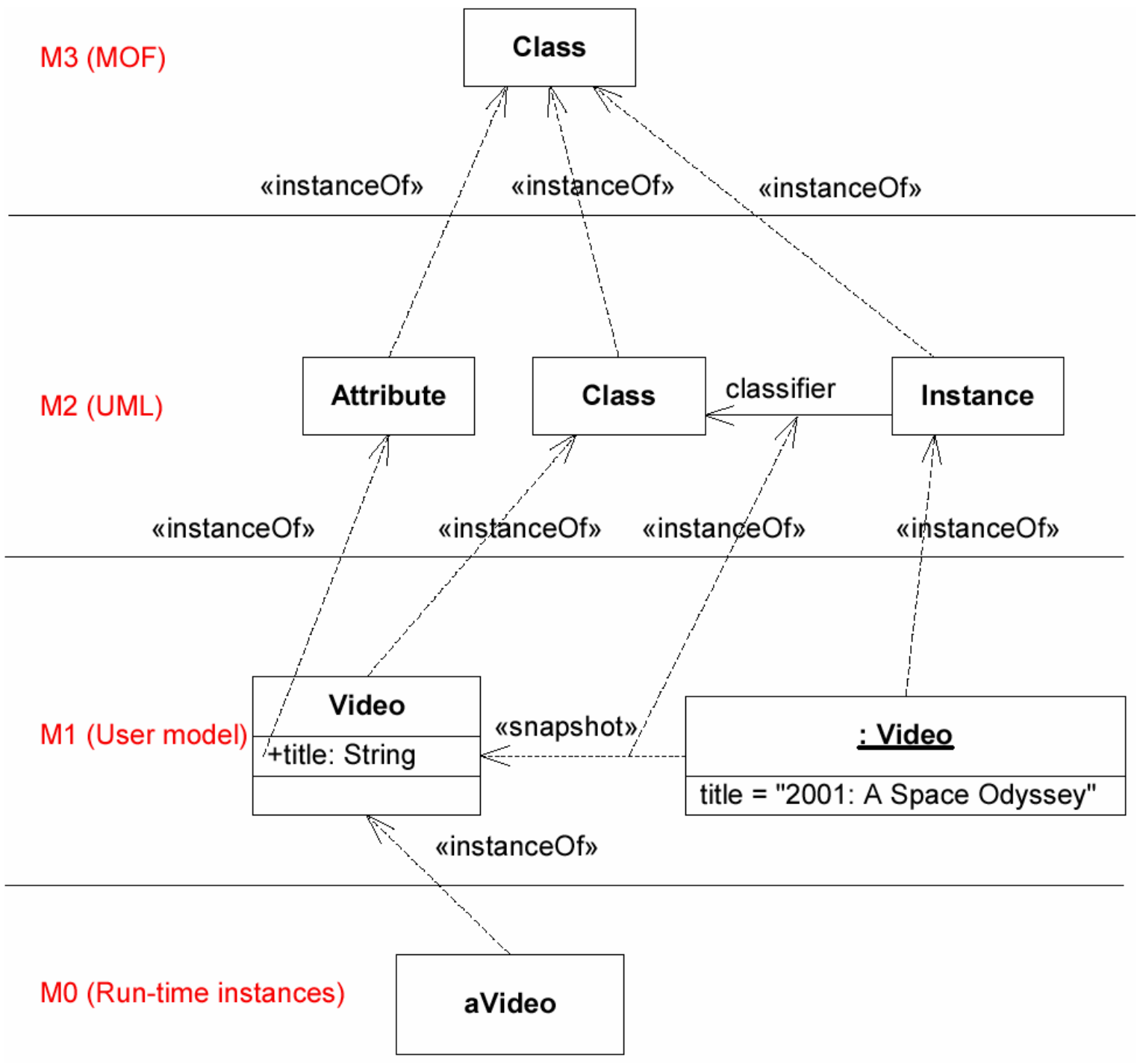

*Fig. 1 Example of the abstraction layers according to MOF*

In Fig. 1 diagram existence of the real life item – video cassette of the movie "2001: A Space Odissey" is described in MOF. We will analyze it from the bottom to the top of the figure:

- M0 layer is considered "real world", e.g. tangible video cassette in the example. If "real life" is *simulated* in the software, (or managed in business process management system), then M0 layer represents "runtime instances" – digital objects – their data structures and methods in the memory. It is easy to understand, that to simulate "real life" (in specified domain) correctly, there should be one-to-one relation between objects in real world and digital world.
- M1 layer shows model of M0 layer instances (i.e. "runtime instances"). This abstraction layer is called "user models", because in this abstraction layer we draw different models of different "business domains". In the example we model video collection. And we see, that there is an object with "type" (class in UML terms) *Video*, with String type property called *title*. And then fact, that there is cassette with title "2001: A Space Odissey" in the model can be shown as "snapshot" of the *Video* class. Formally this model can be read in a following way: there is a collection (class) called *Video,* where there is always particular item (object of this class) called "2001: A Space Odissey". Word "always" is particularly important, because this diagram shows so called "static structure" or things, which are always true. So, if this M1 model is true, in this particular business domain there is *always* cassette "2001: A Space Odissey" (and one cannot remove it from the software).
- According to MOF, concrete "user model" of the "business domain" is an instance of some metamodel (M2 layer). Concrete models can be instantiated from different metamodels, but, according to UML standard, each model is instance of the UML *Class*. In other words, UML standardizes (one of possible) metamodels (UML superstructure [11]). If one uses UML and look at example as a Class diagram, then user class *Video* is instance of the metaclass *Class.* Attribute *title* is instance of the metaclass *Attribute*, but instance *:Video* is instance of the metaclass *Instance.*
- All metaclasses *Class, Attribute, Instance* are instances of the particular metametamodel (M3 layer). Again, there could be different metametamodels, but if UML is used in modeling, then metametamodel is defined in MOF (and UML infrastructure [12]) standard.

Reading diagram form top to bottom, one can say that:

- Model in the M3 layer defines, how one can create valid instances of the M3, or M2 models (metamodels). E.g. UML language is valid instance of the MOF.
- Model in the M2 layer defines, how one can create valid instances of the M2 layer, or "user" models. E.g. Video collection model is valid instance of the UML class diagram.
- Model in the M1 layer defines, how one create runtime objects (instances) in the software.

In theory models of the metametamodels could be analyzed and metametametamodels could be introduced, but in UML interesting feature of the discrete modeling is used. When one look at discrete models models in general, one see "associated" "things" or *Classes* and *Associations*. And there is no general difference if these models describe real things, or models, or even models-of-the-models. Therefore in MOF it is postulated, that there is no more abstract layer as M3, and even more abstract metamodels "*reflect*" the same properties as M3 layer. In other words, models of the metametamodels will be also valid metametamodels, according to the MOF.

Therefore name of classes (in UML) and metaclasses (MOF) is the same – *Class*, because MOF, similarly to other reflective languages (simple example of which is BNF [13]), *can describe rules and semantics of language using MOF language itself.* (Actually MOF, to describe itself, uses *subset* of the UML class diagram [12]).
Other example of MOF approach is following. If we describe native language using MOF approach, then each particular paper (e.g. this paper) is model of the real world (M1). Of course, the same idea could be described using different model (i.e. other language). Then metamodel (M2) describes different languages (e.g. Latvian, English) and its context free syntax (i.e. concepts of the language without grammar). Metametamodel describes means of the language, or abstract semantics and grammar of the language syntax (how words, sentences, questions, tenses, etc. are formed).
To describe MOF approach in relation to other models, Fig. 2 shows how different metamodeling approaches to the real life phenomenon can be used. In the left side of the picture MOF formalism is shown, but in right side – ontology using OWL [14] language:

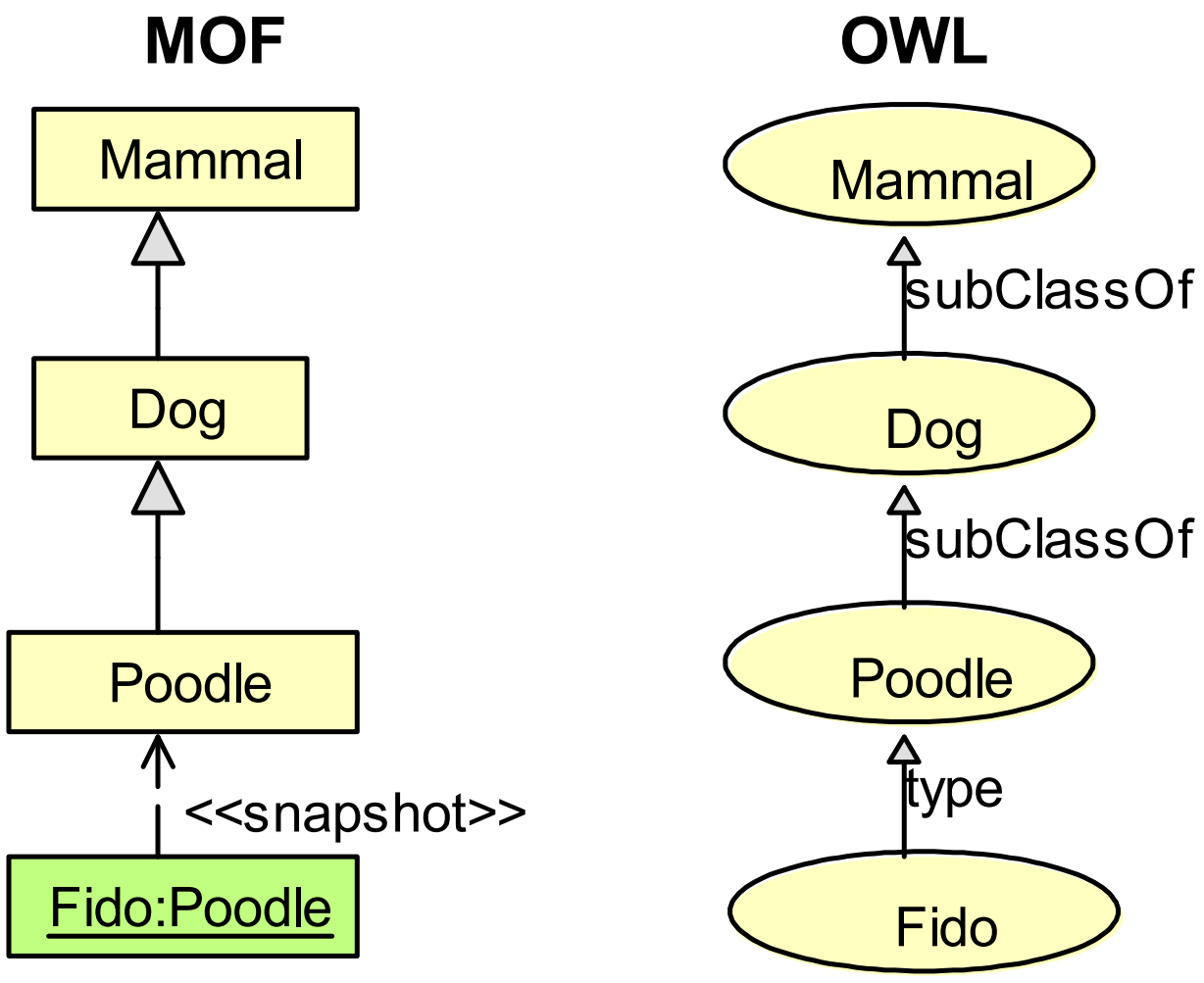

*Fig. 2 Differences between MOF and ontology*

Lets assume, there is real exemplar of Mammal class, Dog specie and Poodle breed (it is *not* exactly right for biology classification) with name *Fido*. According to MOF, it could be shown as concrete "snapshot" of the *Poodle* class, which is specialization of the *Dog* class, which is specialization of the Mammal class. Note that all these facts can be shown *in the same abstraction layer* – M1. <<*snapshot*>> association shows that *Fido* is instance of the same abstraction layer M1. In simple terms, one can see that everything in this model is on the paper, an it *represents* real life but *is not real life* itself. So *Fido* is not M0 instance (real object), but it *Fido representation* in M1 layer.
(It would be M0 layer, if we would create program (e.g. game) with runtime object *Fido*, which would have properties prescribed in the M1 model).
In ontology the same fact is shown differently: *Fido* type is *Poodle*, which is subclass of *Dog,* which is subclass of *Mammal*. In OWL *type* association analog in UML is <<*instance of*>> not <<*snapshot*>> association, because in OWL both – classes and instances are shown. In OWL one diagram represents several abstraction layers and using *type* association one can create arbitrary many abstraction layers.
In this paper MOF formalism is used everywhere. Models represents things of the real (or simulated) life, e.g. structure of the enterprise its resources and different properties of them. Metamodels describe different process modeling languages (syntax of the business models). Metametamodels describe general features of different metamodels (syntax of the business modeling languages).

In this paper important feature of metamodels is used – the same phenomenon of the real (or simulated) life can be described instantiating models from different metamodels. Analyzing different metamodels following equivalence criteria can be enumerated:

- Metamodels are *syntactically equivalent* if any instance set from this metamodel is also valid instance set for another metamodel, including class, association and attribute names. So for syntactically equivalent metamodels all concrete (non-abstract) classes have the same attributes and association (including inherited ones). This feature of equivalence is used to "lfatten" models, by removing unnecessary abstract classes from the inheritance tree for concrete subdomain of investigated languages. Syntactic equivalence of metamodels is formally provable.
- Metamodels are *semantically equivalent* if instance set of one metamodel is also valid instance set of other metamodel in such way that: metamodels have only different multiplicities for associations (constraints) and classes and/or associations have different names. Practically semantically equivalent metamodels have different constraint rules and name the same things in different names. Semantic equivalence can be formally proven, if terms in different metamodels can be mapped from one model to another.
- If two metamodels are not either syntactically or semantically equivalent, but their instance sets describe the same domain of the real life with comparable precision, then such metamodels can be called as *semantically similar*. Semantically similar are different object-oriented programming languages e.g. Java and C#. Also native languages, such as Latvian and English, are semantically similar, because both of them are comparable in their expressiveness. But concrete sentences from both languages are completely different because they do not share common abstract syntax. As semantic similarity is not formally provable, semantic similarity can be shown only on specific examples they do not imply that models are semantically equivalent in general.

Relying on model equivalence and similarity they can compared each to other (i.e. their terms can be mapped), or they can be *transformed* – changed to different models with merges and substitutions.

Syntactically and semantically equivalent models can be transformed automatically using e.g. model mapping approach.

Comparison of semantically similar models is not so straightforward, but for specified sub-domains semantically similar models can be compared using model transformation approach.

# 2 Modeling Business

## 2.1 Business Metamodels

Business process modeling (BPM) began as part of the business process reengineering movement in the 1990s. Initially, business models were conceptual and were used in place of software specifications where the diagrams were clearer than text. To ensure unambiguous understanding of models, the first modeling languages were introduced, and modeling tools were developed. One of the first modeling languages was defined in the Zachman framework [23], which was implemented in the Popkin Software System Architect tool. Other modeling languages also were implemented in tools, for example, GRAPES BM in the GRADE tool [80], ARIS in the ARIS tool [34], and other languages, such as IDEF 3 and UML 1.0 activity graphs.

As modeling languages became more concrete, tools allowed validation and simulation (e.g., ARIS, System Architect [35], GRADE), and specific modeling languages for workflow systems appeared (FileNet, MQ Workflow). Thus, models shifted from software specification to executable programs, and execution semantics of modeling languages became an important issue.

In 2002, when this research was started, business modeling was a broad field. The term BPM (business process *management*) was used to emphasize both the process analysis and execution aspects. The necessity to execute process models required strict requirements for modeling language semantics. Many process modeling languages and modeling frameworks explained business modeling in their own way; therefore, for integrated processing of business processes, a common approach was necessary.

The goal of this research was development of a unified modeling methodology based on a new approach using metamodeling as in UML [11]. As the most popular frameworks were developed before the appearance of UML (e.g., Zachman, ARIS), they were not developed as metamodels. Therefore, analysis of existing frameworks and development of a new, harmonized and comprehensive metamodel was necessary. Using a common metamodel, modeling languages using similar concepts ("conceptually similar" languages) can be shown as specific views of a harmonized metamodel.

## 2.2 Main Business Concepts and their Relationships

To describe a business process, two views are required. One view describes the business process itself: what actions under what circumstances are executed by performers in what order. Another view is required for process explanation: what is important for process integration in one or several enterprises. It shows the process environment: inputs (suppliers) and outputs (goods and services for customers), process relationships with other processes, as well as the meaning goals of processes, and quality measures.

An investigation of existing frameworks based on metamodels and other approaches [23,24,22,25] revealed that they were incomplete; none is sufficiently general and comprehensive. Therefore, the author offers a new business process metamodel, in which the most popular frameworks are combined. The concepts coming from different approaches are harmonized into a single metamodel to show only the essential concepts that are common in all analyzed business management methodologies.

Fig. 3 shows the business process environment metamodel developed by the author. It conforms to several of the world's leading business process management standards (e.g., the value added chain and the ISO quality standard [26]). The metamodel is similar to the Business Motivation Model developed by OMG [51]. The Business Rules Group started development of this standard in 2000, but the metamodel was developed in 2005, when it was adopted by OMG, and is still in draft form. It should be noted that, for business process resources, the input and output in the author's research is even more detailed than is required by the OMG standard. In 2006, a similar model was created as a business ontology at the University of Lausanne [56].

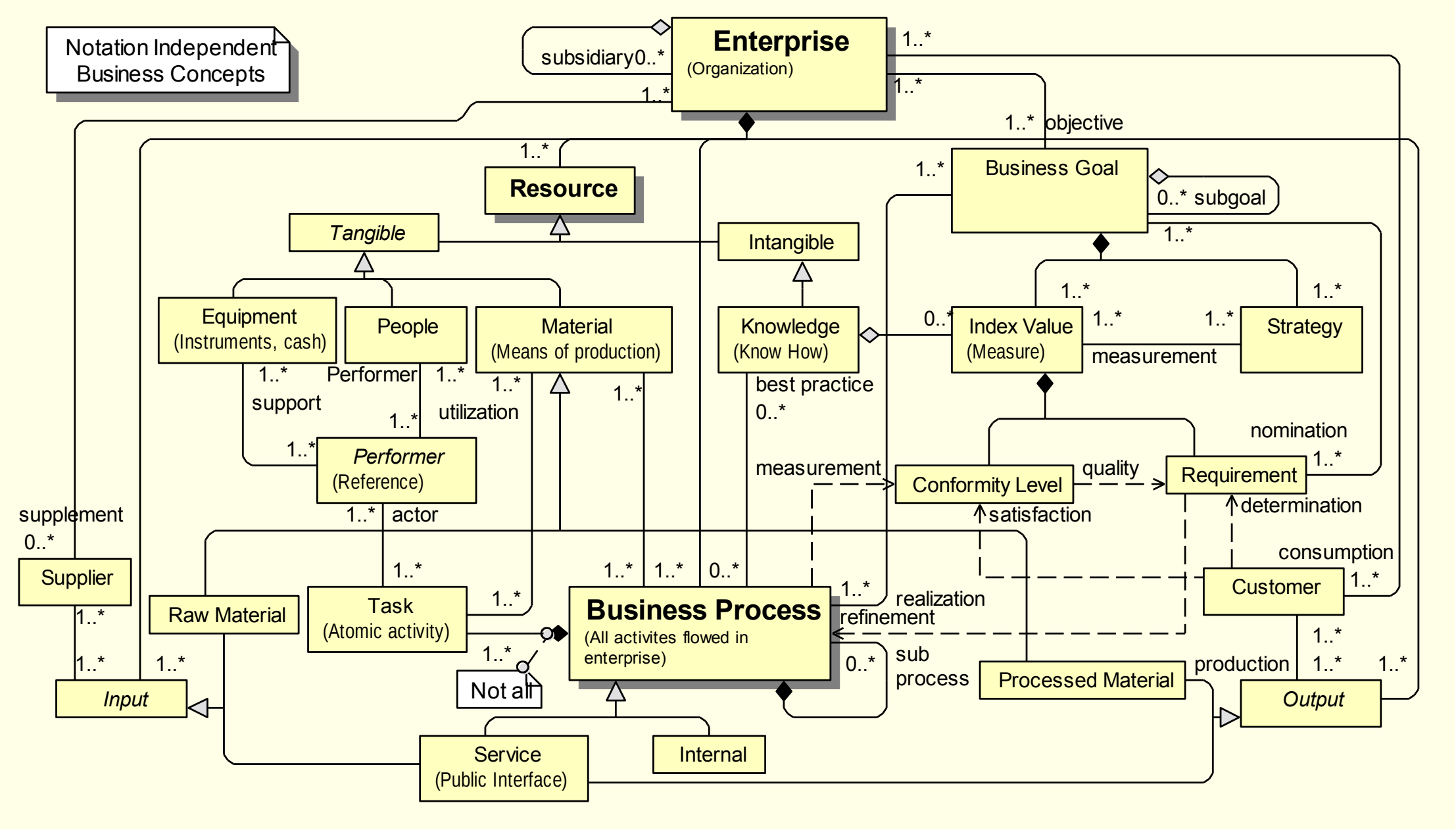

*Fig. 3 Business process environment metamodel*

Fig. 4 shows the business process elements required for process definition and execution. The main concept for all process modeling languages in the *Process* is the *Task*, which represents one atomic function. In different modeling languages it is named differently: in GRAPES BM and BPMN, it is named *Task,* in UML 1, *Action State,* in UML 2, *Action,* in IDEF 3, *Unit of Behavior.* A *Task* is performed by a *Performer*. A *Transition* determines the sequence in which several *Tasks* are executed. *Task* and *Transition* are the main concepts in all business modeling languages.

Most languages also have *ControlElements*, which determine process branching. A *Decision* represents the start of branching; a *Fork*, the start of parallel flows; a *Merge,* the unification of branches; a *Join*, the unification of concurrent threads. Explicit *Start* and *End* points also can be shown. However, control elements and transition connections to the control elements may differ considerably in different modeling languages. Therefore, to unambiguously map specific control elements in the analyzed UML 2.0 AD and GRAPES BM languages, additional transition subclasses are introduced: *SimpleTransition*, *Incoming* transition and *Outgoing* transition. Usage of these subclasses for explicit concept mapping is described in Section 2.3.

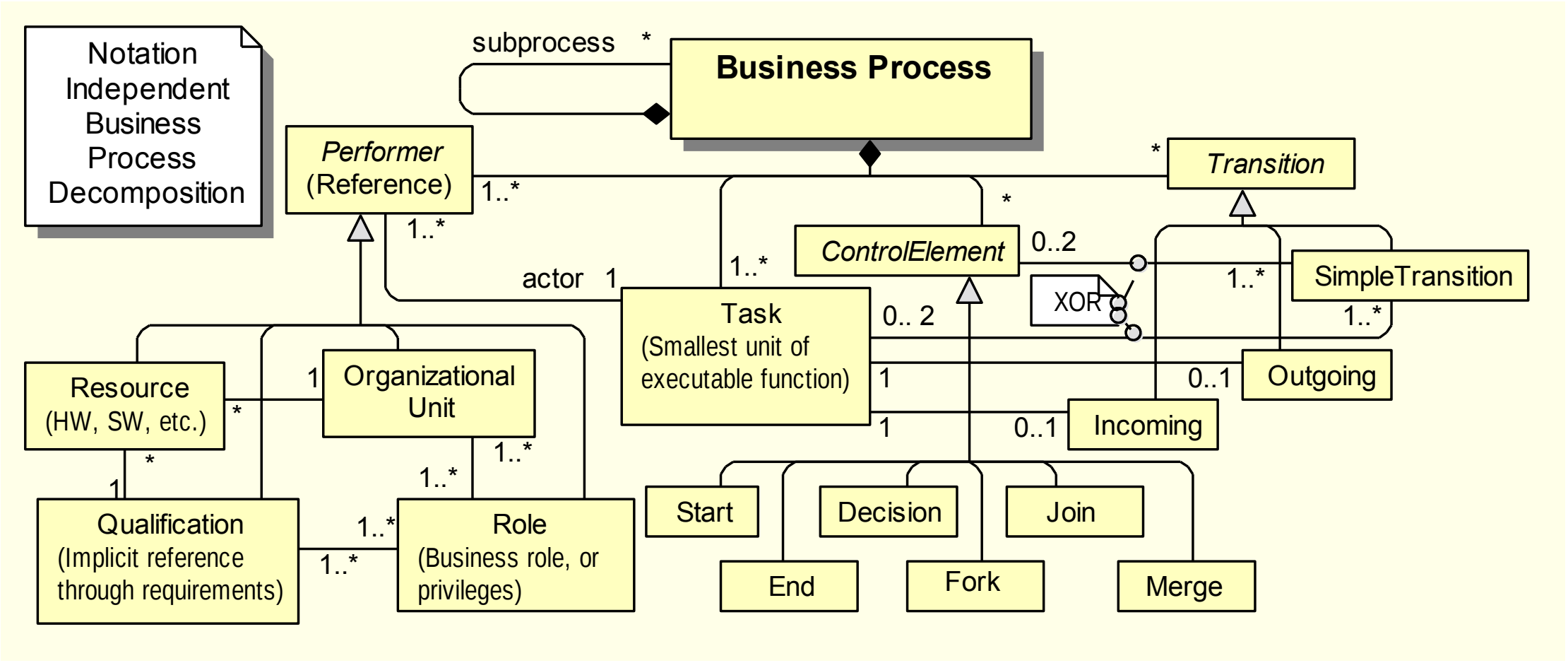

*Fig. 4 Business process metamodel containing classes*

This business process metamodel with classes is similar to that of the OMG process definition metamodel standard draft, which was developed later [50]. It should be noted that part of the metamodel for task performers in the author's metamodel is more detailed than in the standard draft proposed by OMG.

A harmonized or "notation independent" metamodel of the business process environment and its elements (Fig. 3, Fig. 4) is used as a "canonical form" of the business process metamodel in all further research (Chapters 3, 4 and 5). Though the developed metamodel is small, it is widely applicable because it shows the main business concepts of any enterprise. Therefore, it conforms well to standards that were developed later.

## 2.3 Concept Mapping in Different Notations

Using a "notation independent" metamodel, the author has developed a method by which "semantically similar" modeling languages can be defined as specific views of the metamodel. The languages used for this example are UML 2.0 AD (which was only a draft of the standard version at that time) and GRAPES BM. In Fig. 5, the same business process is shown in these two notations, at the same time showing the differences between these languages, e.g., guard nodes in the GRAPES BM [80] language are shown as guard conditions for edges in the UML AD [11] language:

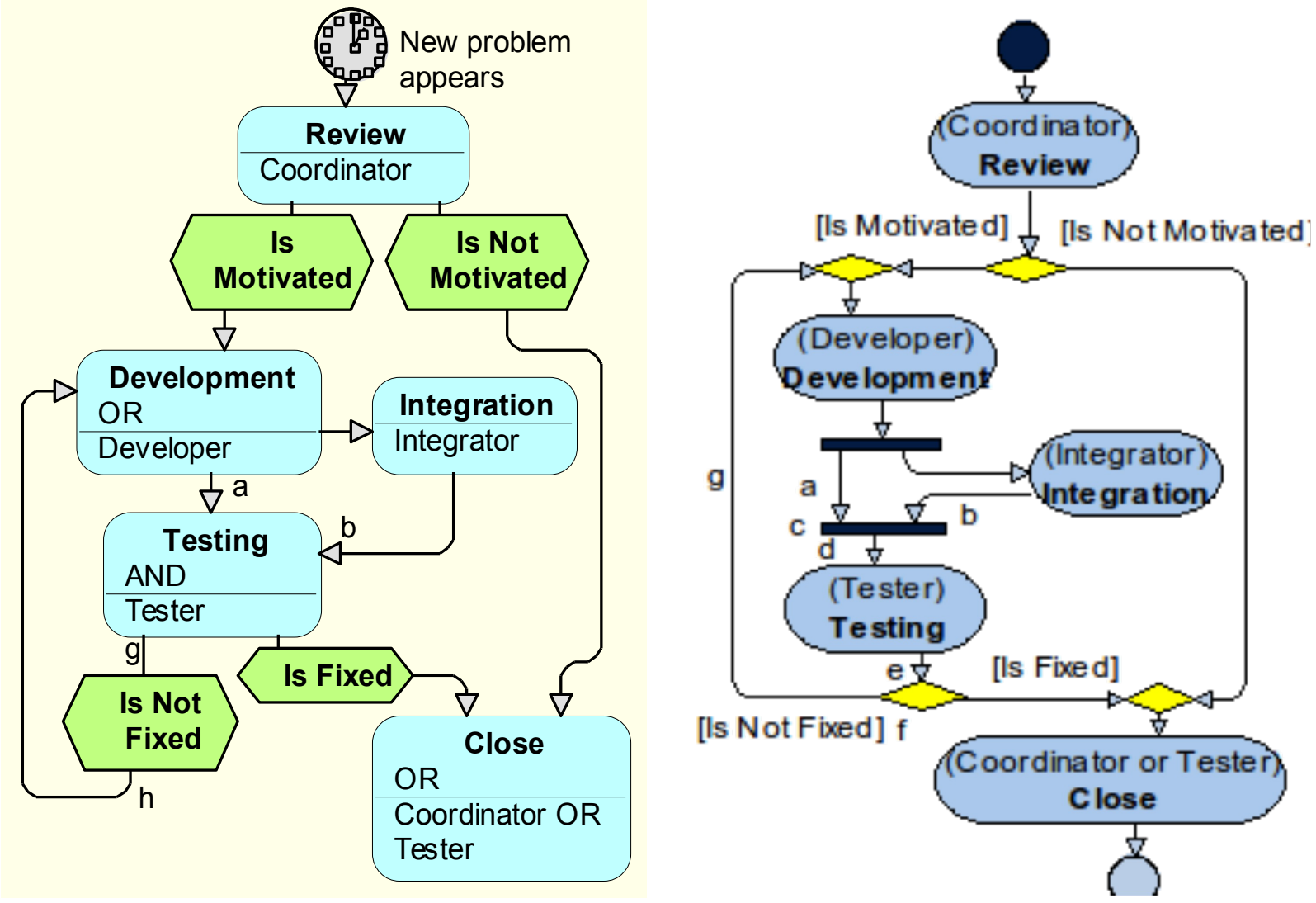

*Fig. 5 A business process as GRADE BM and UML AD*

The general concept mapping schema is shown in Fig. 6. The top part of the figure shows concepts necessary for business modeling (*Domain concepts*). Concepts in *Notation A* are mapped to concepts in *Notation B* using *Notation independent* concepts as mediators. In many cases, relations between language concepts are many-to-many, and are thus not sufficiently clear and traceable. To "normalize" such mappings and make them unambiguous, "intermediate concepts" are introduced, which decouple many-to-many relations between language concepts as a pair of one-to-many relations between language and intermediate concepts.
When mapping is explicitly defined using intermediate concepts, concept instances from one notation are converted to instances in another notation, as shown at the bottom of the figure. Dependency lines show <<instance of>> relations between classes and their instances. Even though, in general, case instance relations are many-to-many, for particular instances this relation is usually one-to-many, as shown in Fig. 7.

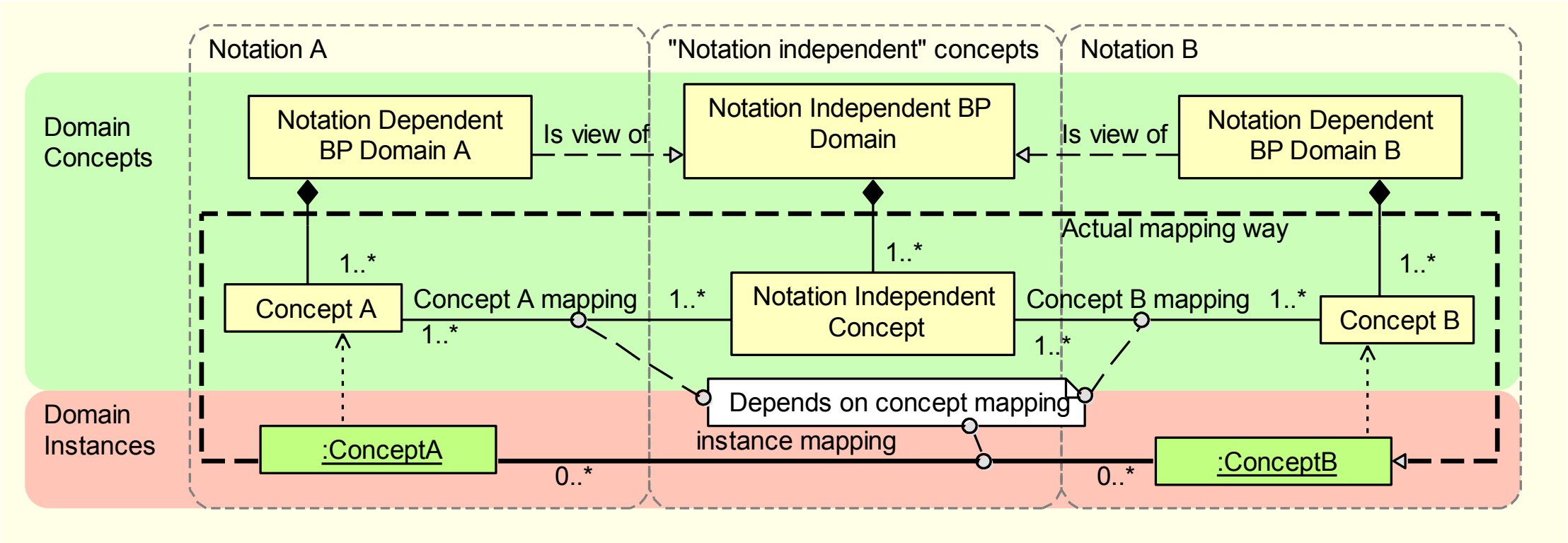

*Fig. 6 The general scheme of metamodel mapping*

Fig. 7 shows a fragment of the defined mapping between the GRADE BM and UML AD languages. The top part of the figure shows the domain concepts layer with GRADE metamodel classes (top left area), the UML AD metamodel (top right) classes, and the independent domain (top middle) classes. General many-to-many relations between GRAPES BM and UML AD for guard nodes and edges are decoupled using several one-to-many mappings between language concepts and intermediate concepts (*Transition* subclasses: *Incoming*, *Outgoing* and *SimpleTransition*). The bottom section of Fig. 7 shows specific instances of these classes based on the business process example shown in Fig. 5. As shown in Fig. 7, a relation for a particular instance for both languages appears as one-to-many, so they are easily traceable.

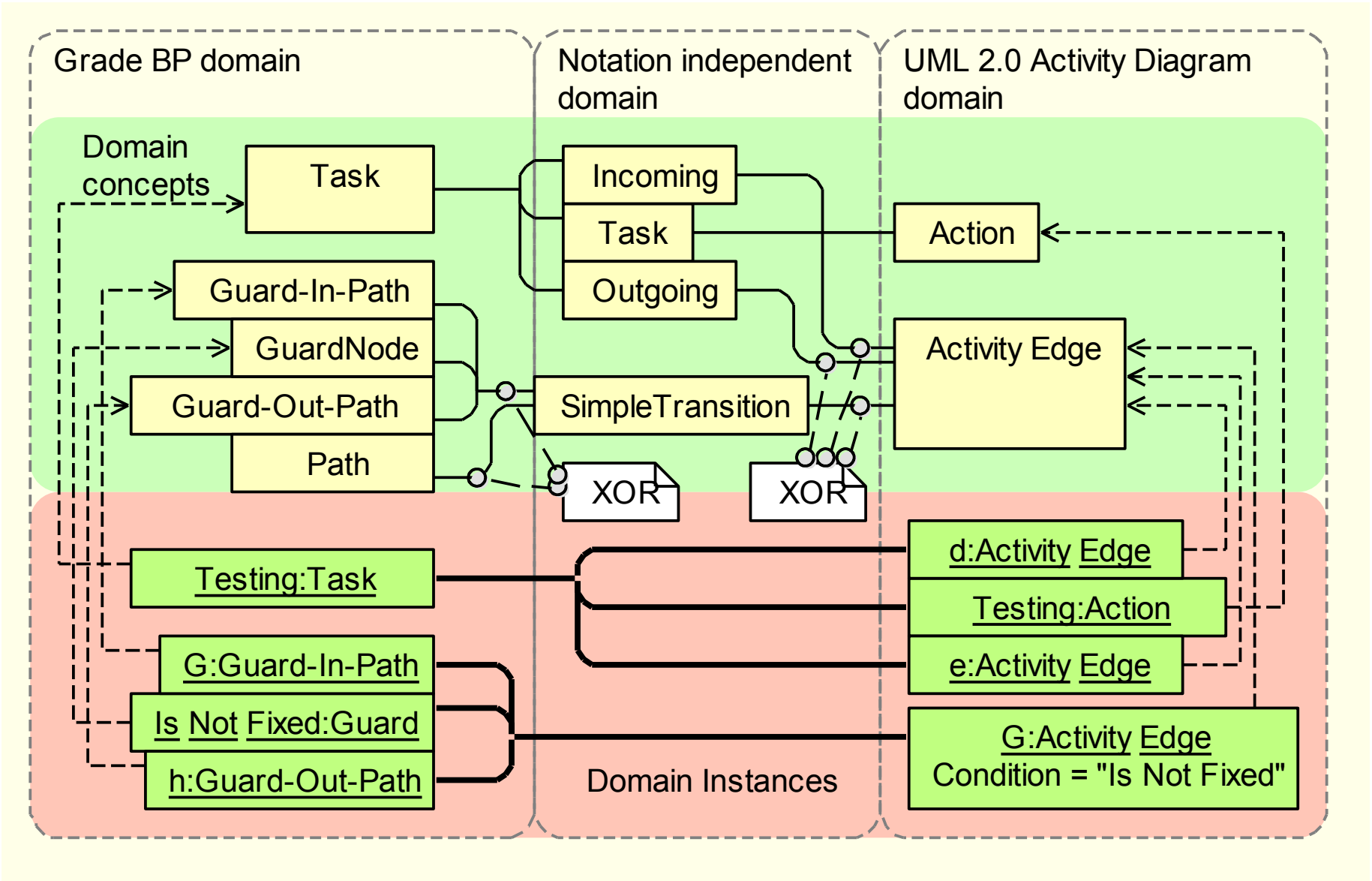

*Fig. 7 Fragment of a mapping definition and class instances*

As users are actually interested in a graphical representation of these two notations, the last step is creation of corresponding graphical elements (boxes and lines) in the diagram. Because domain elements are usually linked to a graphical representation as one-to-one, it is a simple task. As shown in the Fig. 7 mapping fragment, even for "semantically similar" languages technical elements can differ considerably, and this complicates concept mapping.

At the moment, a concept mapping technique is used by the OMG Domain task force group for merging the UML 2.0 AD and BPMN 1.0 languages [57]. Because UML AD and BPMN have even more differences than UML and GRAPES BM (e.g., AD does not have a business process equivalent to that of BPMN, which contains several processes, but BPMN does not have an equivalent to AD object flows), concept mapping either cannot be shown unambiguously, or many intermediate concepts must be introduced. However, for cases when "semantically similar" languages are also "technically similar," concept mapping is good shorthand. Recently, this approach has shown some success, e.g., for mappings between the BPMN, BPEL, XLANG and WSFL languages [59].

Subsequent research (described in Chapter 5) showed that complicated mappings can be effectively resolved using model transformations. In this case, mapping associations are doubly useful: in definition, they demonstrate the general mapping schema (though ambiguous), but for specific instances, they show the transformation result, pointing to the source instance for each target instance. This is necessary for model transformation traceability.

# 3 Business Process Measures

As the business world becomes more global and competitive, business processes are becoming more geographically distributed. To estimate overall process efficiency, it is essential to recognize costs, time and other parameters of each business process step. Business process efficiency measures are crucial to determine total costs of ownership and forecast the return on investments. Monitoring and measures are a ubiquitous element in workflow systems [46].

Many quality and business process management methodologies use numeric methods to measure the weaknesses and strengths of a business [30,31,32,33,26]. These methodologies are supported by several tools [34,35,36,37]; however, they provide the "best of the breed" methodology only for narrow areas, and cannot support several methodologies simultaneously. Measure definition and value calculation, especially for aggregations (e.g., sums, averages, minimums and maximums) are not trivial tasks using existing simulation tools and workflow systems. Advanced technical skills are required.

The goal of the research was to develop a business process measure framework that uses algorithms and constraints to measure business process elements and their aggregations in a way that provides practical data. This framework should support business measures that are defined in the business model itself. It would allow business analysts to work with measures using familiar business terminology and avoid having to deal with technical issues.

This research shows that such a framework can be developed using a metamodeling approach. The process measurement language is developed using a profile from the UML AD [11], but a measurement processing framework is provided through heavyweight extension of the UML metametamodel [12].

The approach is described by presenting a business process measures example, and then providing an abstract measure definition framework.

## 3.1 Business Process Model

To demonstrate the principles of measure calculations, a simple business process example is used. It is defined using a UML AD [11]. Fig. 8 shows the business process for a shop that delivers pizzas to residential customers. The *Sell Pizzas* business process (activity in the AD) contains actions (rounded rectangles). The organizational structure is not shown separately; it can be assumed from the process description. Organizational units are shown in swimlanes, and their performers (positions/roles and resources) are shown in action compartments with parentheses. Performers pointing to positions/roles (*People*) are active: they perform manual operations. *Resources* are passive. They are necessary to perform an action and can be busy or spent in operation. The object flow is shown using flow nodes (rectangles) and data stores as parallelograms. Process measures are shown in notes (rectangles with bent corners), which are attached with dashed lines to measurable objects.

According to MOF [16], this business process model is an abstraction of all execution instances of the real business process; therefore, it conforms to the M1 layer.

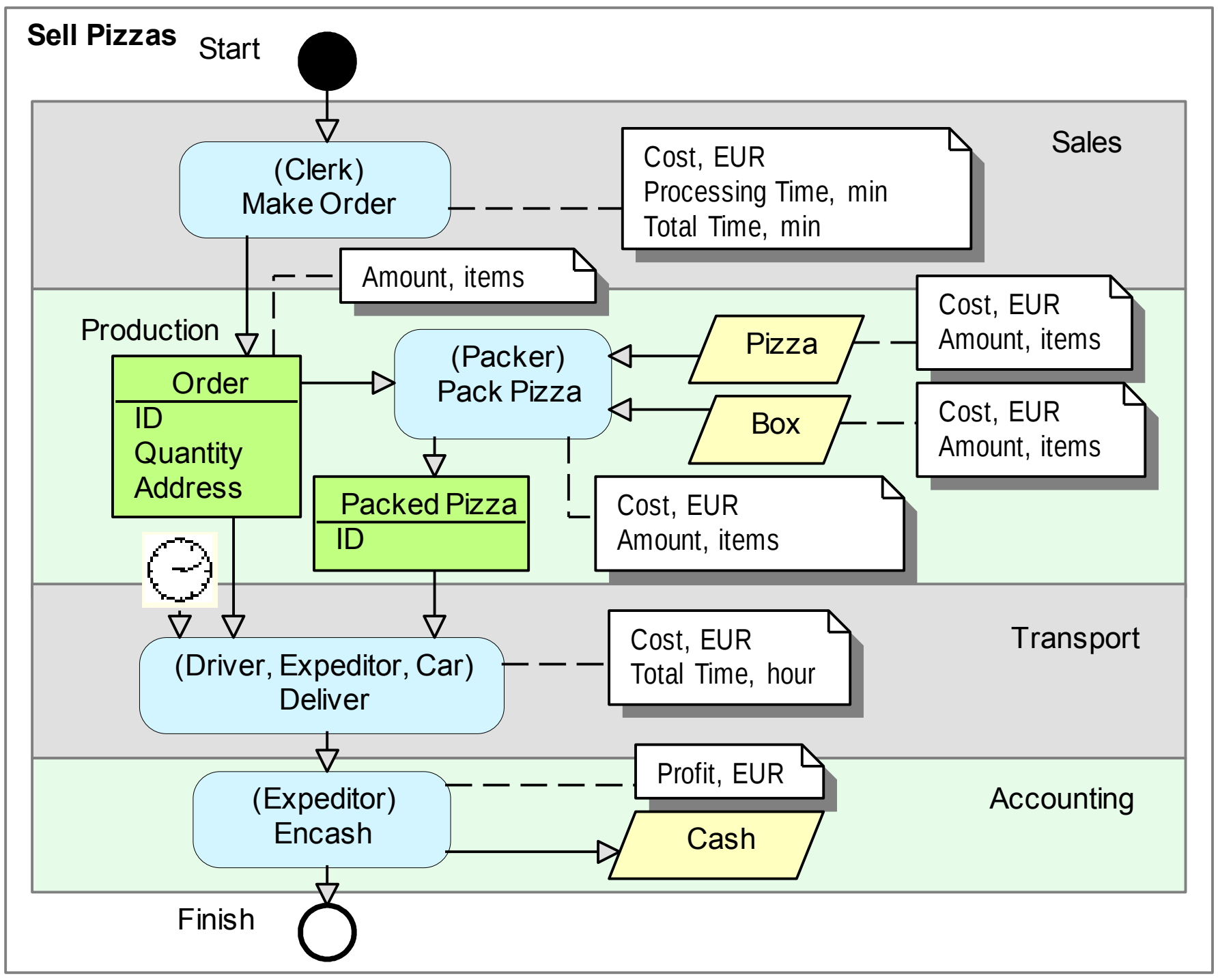

*Fig. 8 Sell Pizzas business process as an Activity Diagram* (*M1*)

The *Make Order* action invokes a subprocess (activity), which is shown in Fig. 9.

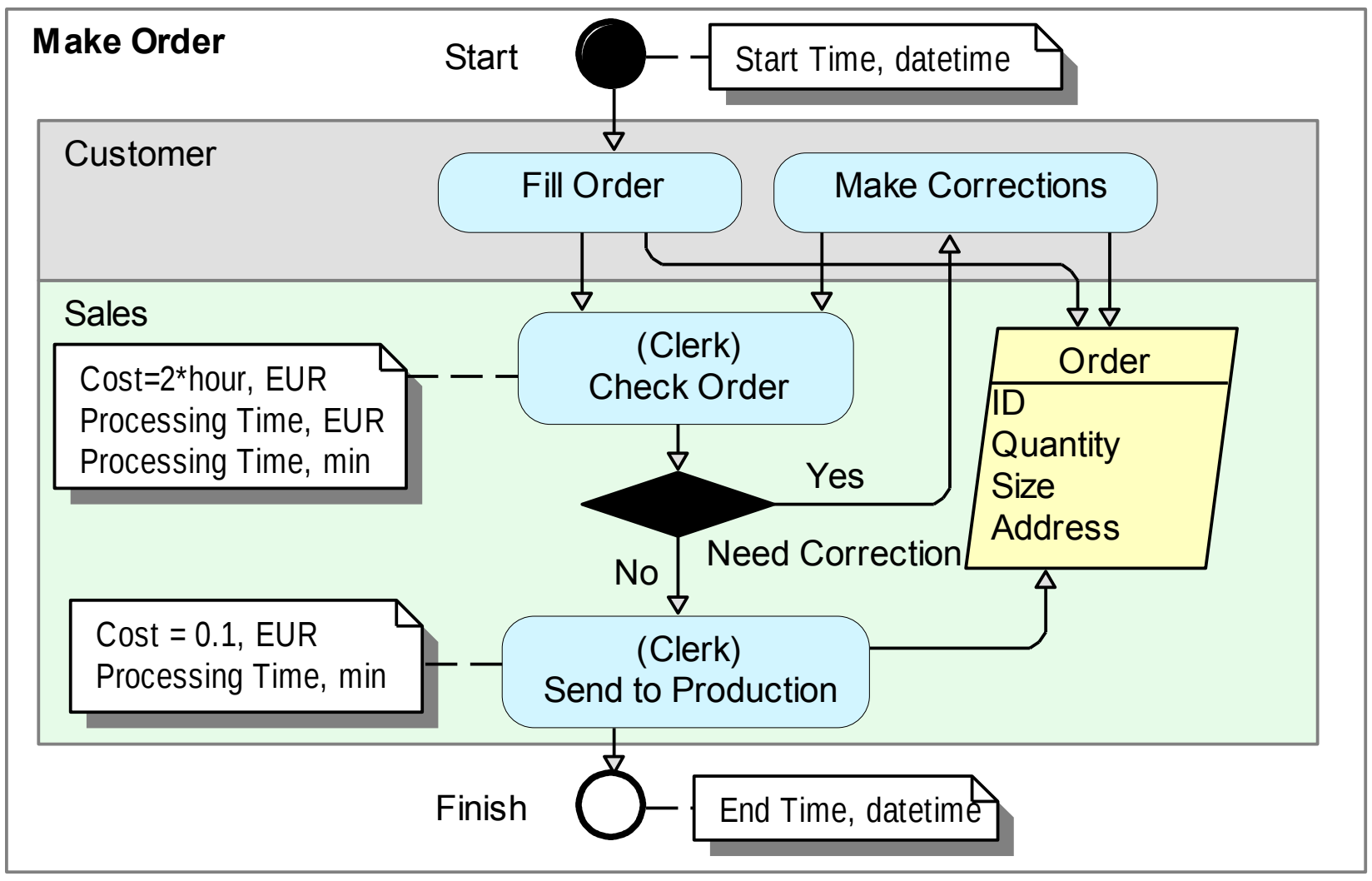

*Fig. 9 Make Order business process (M1)*

To support business measures, the UML AD is extended according to the UML standard with a measure definition profile. Process measures actually are attributes of the metaclass stereotype, which according to UML can be shown as notes. To improve readability, several object measures are joined in one note. Each measure declaration has the following syntax: Name[=declaration],Unit (e.g., *Cost=2*hour, EUR*). A measure linked to a process element means that the given measure must be evaluated during process simulation or execution.

This example shows that for process measure definition (measure specification), it is possible to use a UML AD extended with class stereotypes for measures, similar to existing simulation systems (e.g., ARIS [34]).

## 3.2 Measure Aggregation Sample Model

Although it is possible to use a UML profile for an AD to define business process measures, that is not sufficient. To show exactly how numeric values of measures are processed in process execution time and how default operations are performed, additional classes and associations are necessary. Therefore, the UML AD is only an external specification or interface of the framework, but its implementation is much more complicated. Within the framework, the AD should be transformed to a specific "internal model," in which, in addition to the UML AD measure profile, the diagram is extended with new classes and associations (a heavyweight extension). Because the MOLA modeling language [82] was not developed at that time, the author has shown the AD transformations using the possibilities supported by UML class diagrams using specific syntax.

The diagram in Fig. 10 shows part of the processes defined in Fig. 8 and Fig. 9 as an internal model of a business process measure calculation framework. This model is "execution independent" because, regardless of the implementation, it shows which elements, associations and calculations are required to calculate values of the defined measures. Formally, such a model should be shown using two diagrams: a class diagram in which classes describe all instances appearing in process simulation or execution (M0), and an instance diagram of the framework metamodel described later (M2, Fig. 11). Because both diagrams should show dual properties of the same object, for presentation economy they are joined in one class diagram, in which dual "instance classes" are shown with a specific syntax. To show that a class in M1 is also an instance of a more abstract class in M2, this "instance class" is shown using stereotypes, that is, the name of the more abstract class (M2) is shown as a stereotype for a specific abstraction level (M1) "instances class."

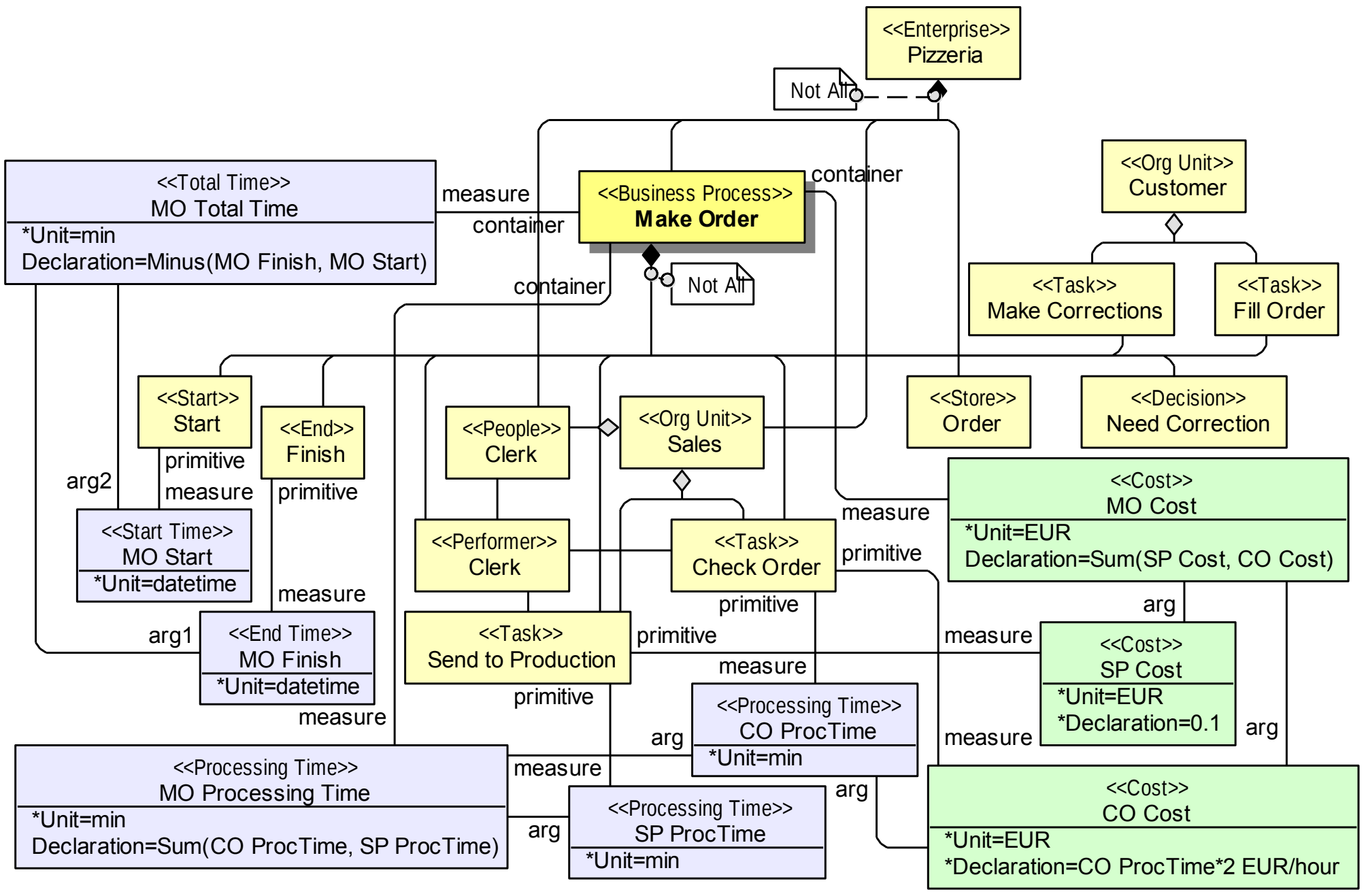

*Fig. 10 Measure aggregation model example* (*M1*)

Light classes are basically stereotypes of the UML AD. As a result of the metamodel heavyweight extension, the AD has additional aggregation associations (e.g., the *Sales* aggregation). Process measures (different shades represent different measure types) linked to process elements with *primitive/container-measure* associations are AD stereotype attributes, which are shown as "instance classes." For model readability, the measures are assigned names consisting of the first characters of the linked element name and the measurement type. (The name is not required for processing). If a measure uses another measure, the appropriate association has the role name *arg*.

The measures composition derived from the M2 layer in the M1 layer is shown according to MOF traditions, using class compartments (e.g., *unit=EUR*, *declaration= Minus*(*MO_Finish,_MO_Start*)). Compartments defined explicitly in the business process model in Fig. 8 are shown with an asterisk. Other compartments and appropriate measure associations are derived implicitly, by creating instances of the process definition metamodel. E.g., the *MakeOrder* business process measurement *MOTotalTime* declaration is determined automatically as the difference between the process start, *MOStart,* and end, *MOEnd,* values with appropriate associations.

For model readability, other M2 level class composition instances also in the M1 level are shown as a composition. E.g., *Business Process* and *Enterprise* M2 composition instances are shown as an M1 composition with appropriate classes and their stereotypes.

## 3.3 Business Process Metamodel

To define rules for process element measures definition and aggregation, a specific metamodel (M2) is required to serve as a framework for all metamodel instances or process definition models (M1).

The developed business process measure metamodel is shown in Fig. 11. The figure shows how measurements are linked to each together, and how they are linked to business process elements. In this figure two diagrams are merged also: class diagram of all possible models (M1, Fig. 10) and a specific instance diagram of a more abstract metametamodel (M3, Fig. 12). This business process measure metamodel basically conforms to the business process metamodel (Chapter 2) with the following differences:

- All classes are instances of the extended metametamodel (M3).
- Although light classes conform to UML AD profile stereotypes (e.g., *BusinessProcess* as a stereotype for *Activity*), in addition to the associations defined in UML, by using an extended metametamodel (M3), additional associations are possible.

Associations defined in the UML and additional ones are used to determine which elements are eligible for measure aggregation. For this reason, a completely new binary association could be used. Though using existing UML associations, the number of additional associations is reduced and the metamodel is more readable. Aggregation (composition) associations show that for an appropriate "host" element, instance aggregation of the same type of measures is possible (e.g., sum, average, minimum, maximum) for all child instances.

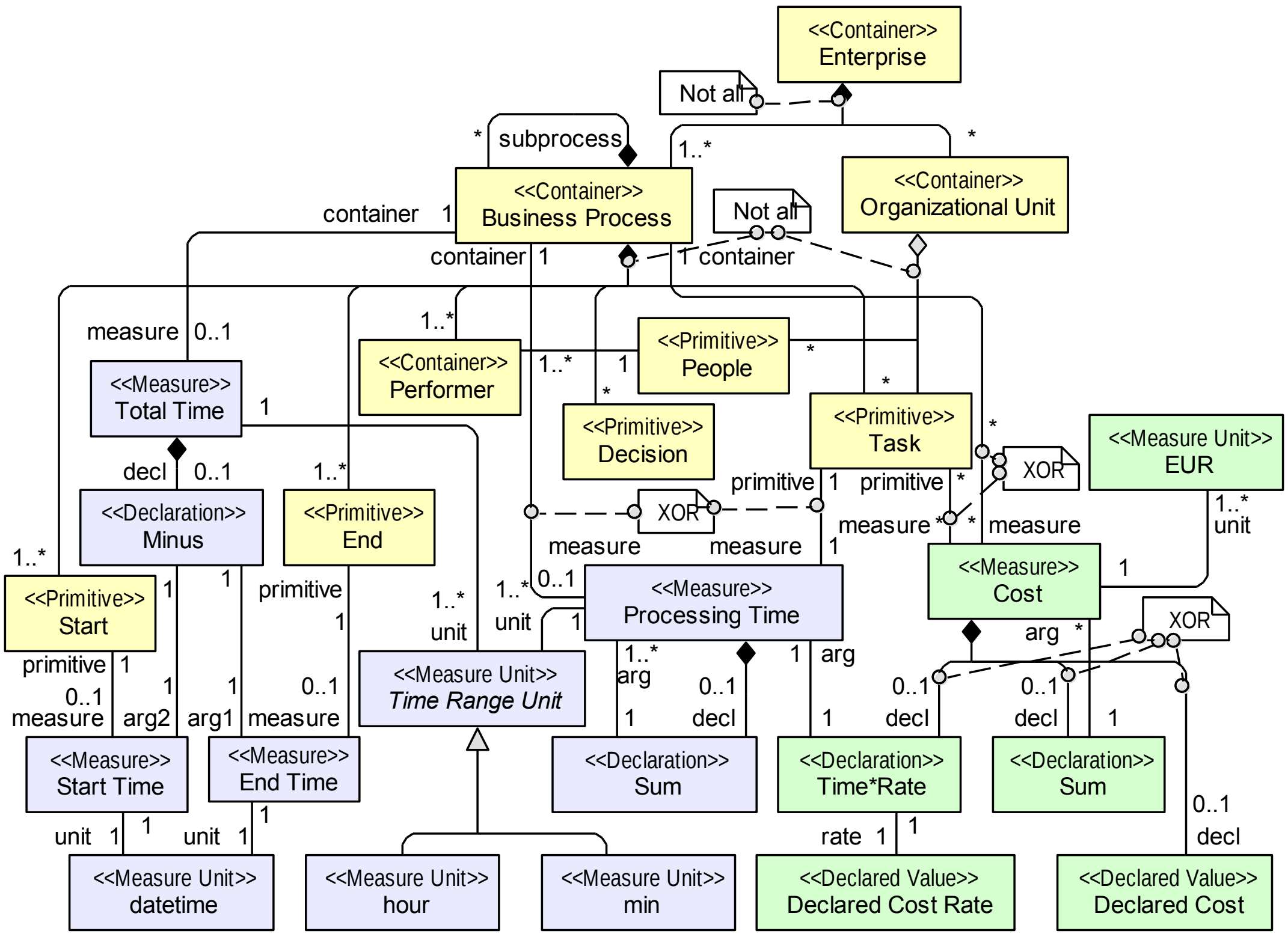

*Fig. 11 Fragment of the measure declaration metamodel* (*M2*)

*<<Measure>>* classes (different shades represent different kinds of measures) are class metaattributes (formally, UML 2.0 properties), which are carried out as separate classes and associations. This diagram also serves two purposes: as a class diagram for the M1 metalayer, and as an instance diagram for the M3 metalayer. Therefore, in accordance with the dual notation, metaattributes appear as classes. These metaattributes show the measures that can be defined for each process element (light classes).

Associations between stereotypes and metaattributes inherit specific role names: *primitive/container-measure*. If a measure declaration has an expression (which can be derived implicitly from the metamodel, or set explicitly in the model), then it is shown as a function argument set with the *<<Declaration>>* stereotype.

Fig. 11 shows possible business process measures. E.g., for a *Business Process* or *Task,* the cost can be determined as specified value (*Declared Cost*), amount of processing time multiplied by the cost rate per time unit (*Processing Time, Declared Cost Rate*) or sum of several costs.

The main value of this metamodel is that it shows in a demonstrative way with classes and associations the possible measures of business process elements and how measures can be declared and aggregated in a way that has practical meaning. By creating models on the basis of this metamodel, defined classes show which measure instances can be added to the model element instances, and associations show the links between measures and methods for automated processing.

## 3.4 Business Measure Metametamodel

In accordance with MOF traditions, the metametamodel (M3) is kept simple, and all the complexity of a specific domain is represented in metamodels (M2). However, when using a general metametamodel to show specifics and the meaning of the metamodel (M2), it is necessary to use many OCL constraints. Therefore, the author uses another approach (which is in line with MOF standards). The UML metametamodel is extended with specialized classes and associations, which clearly show business process measures and their relations.
The developed metametamodel is shown in Fig. 12. The main distinction is that it includes new *BusinessObject* and *Measure* metaclasses as specializations of *Class* from the UML *InfrastructureLibrary::Constructs*. This specialized metametamodel makes it easier to create a metamodel because, as shown in Fig. 11, even a small M2 fragment with a specialized M3 is complicated. Creating an M2 with standard UML MOF and OCL [15] constraints would be much more difficult, and the metamodel would be considerably larger.

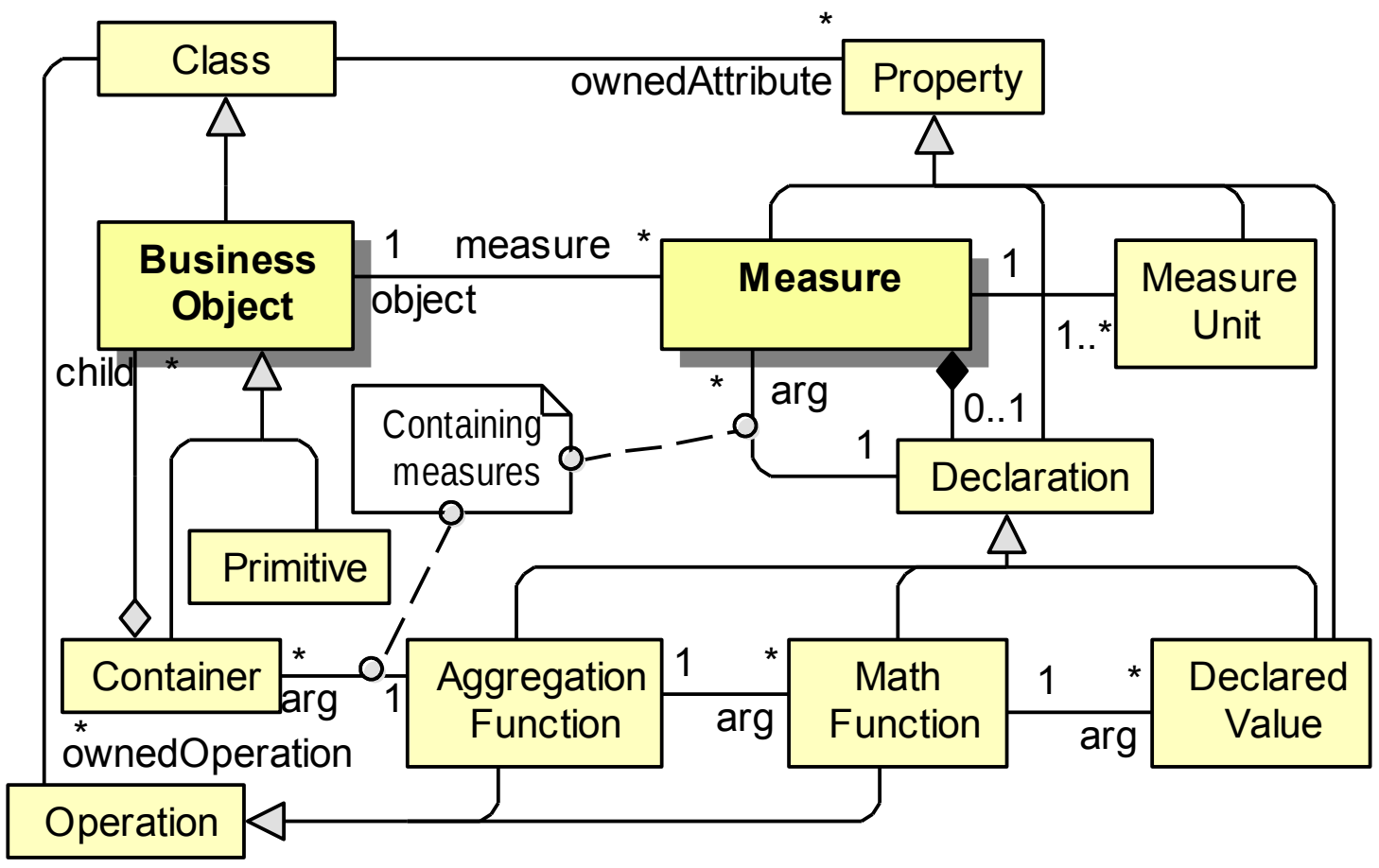

*Fig. 12 Business process measure metametamodel* (*M3*)

In developing the business process measure framework, it is shown how effectively two types of UML extensions can be used. For business process measure definition, a UML AD can be used, extended with a business process measure profile. Profiles are shown as stereotype attributes. In this way, process measure definition compatibility with UML is provided. On the other hand, a measure processing framework can be provided in a clear and effective way, through heavyweight extension of the UML metametamodel. In this way, it is assured that measure constraints and aggregation logic are defined as demonstrative classes and associations between measures, and between measures and measurable elements in the business process metamodel.
Measures are only declarations or definitions of how business objects will be measured. Actual values are obtained only at system run-time. For process definition execution, the possibility of exact semantics dependent only on measure declarations should also be analyzed. In this paper, run-time aspects are discussed only briefly. System run-time aspects (not detailed for measures) are analyzed in further research as shown in Chapter 4.

# 4 Semantics of UML 2.0 Activity Diagram for Business Modeling by Means of Virtual Machine

In 2004, the UML 2.0 standard [11] was in the final adoption stage, and its UML Activity Diagram (AD) was suggested for use in business process modeling. Therefore, exact execution semantics of AD was a topical issue.

Currently de-facto standard for modeling of workflow system behavior is based on finite state automata which, particularly in UML, are described in State diagram. State diagrams are hard to use distributed systems, because they all should be described as some pseudo-state of the global automata. In difference AD show information/data flow between separate parts of the distributed system and therefore are more feasible for more complex workflow systems.

The main goal of this research was to make the AD usable for modeling exact (i.e., executable) business processes. To achieve this, it was necessary to determine the exact execution semantics of ADs and select an appropriate subset of its elements, which was sufficient for business process modeling model validation and simulation. In the research, it was shown that original AD semantics defined for executable models is complicated and not described clearly enough. Therefore, the author proposes a new approach, describing existing AD semantics with an activity diagram virtual machine (ADVM). To describe the semantics, a minimal subset of AD elements was chosen as required for business process definition. It is shown how a formal AD definition model can be translated to a simpler and more convenient execution model, which at the same time conforms to the original execution semantics for the selected subset of AD elements.

Such an approach, which relies on the original AD notation as much as possible in defining semantics, in contrast to absolutely formal algebraic methods (e.g., Petri nets [39]), is more suited for exact analysis of diagram behavior, even if it does not support formal mathematical analysis. In addition, the developed VM can be used as a basis for practical implementation of a simulation tool or workflow management system engine.

## 4.1 Subset of the UML 2.0 Activity Diagram and UML Limitations

Because a number of UML 2.0 AD concepts are actually optimized for development of embedded systems, the author has chosen only those elements that are necessary for business modeling. AD model behavior is determined by tokens, which flow from node to node through edges starting at an activity initial point and ending at the final point. Different tokens can be used for data transfer from one action in the activity to another. Therefore, when determining required elements, only those are chosen that have a direct influence on token movement. In practice, these are actions and control nodes, as well as edges with guards connecting them.

There are few works analyzing the execution semantics of AD control flows alone [38,39,40]. An analysis of the AD data flow separately from the control flow in [41] concludes that AD is not "workflow complete." However, in the author's research analyzing data and control flows in an integrated approach, it is shown that ADs are usable for formal and exact definitions of business processes.

Fig. 13 shows two activity diagrams, which illustrate all chosen elements. The main AD *Process Order* invokes another one: *Make Payment*. The main process starts with the initial node, then the process flows through decision, fork, join and merge nodes and finishes in the activity final node. The *Make Payment* action invokes the subordinated activity, which starts and finishes with activity parameter nodes. All flows (control and object) have pins at their action ends (and also at the initial and final nodes), as required by the selected subset. The advantage for semantics definition of using these explicit pins is that tokens always have a place to "live," much in the same way as places are used in Petri nets.

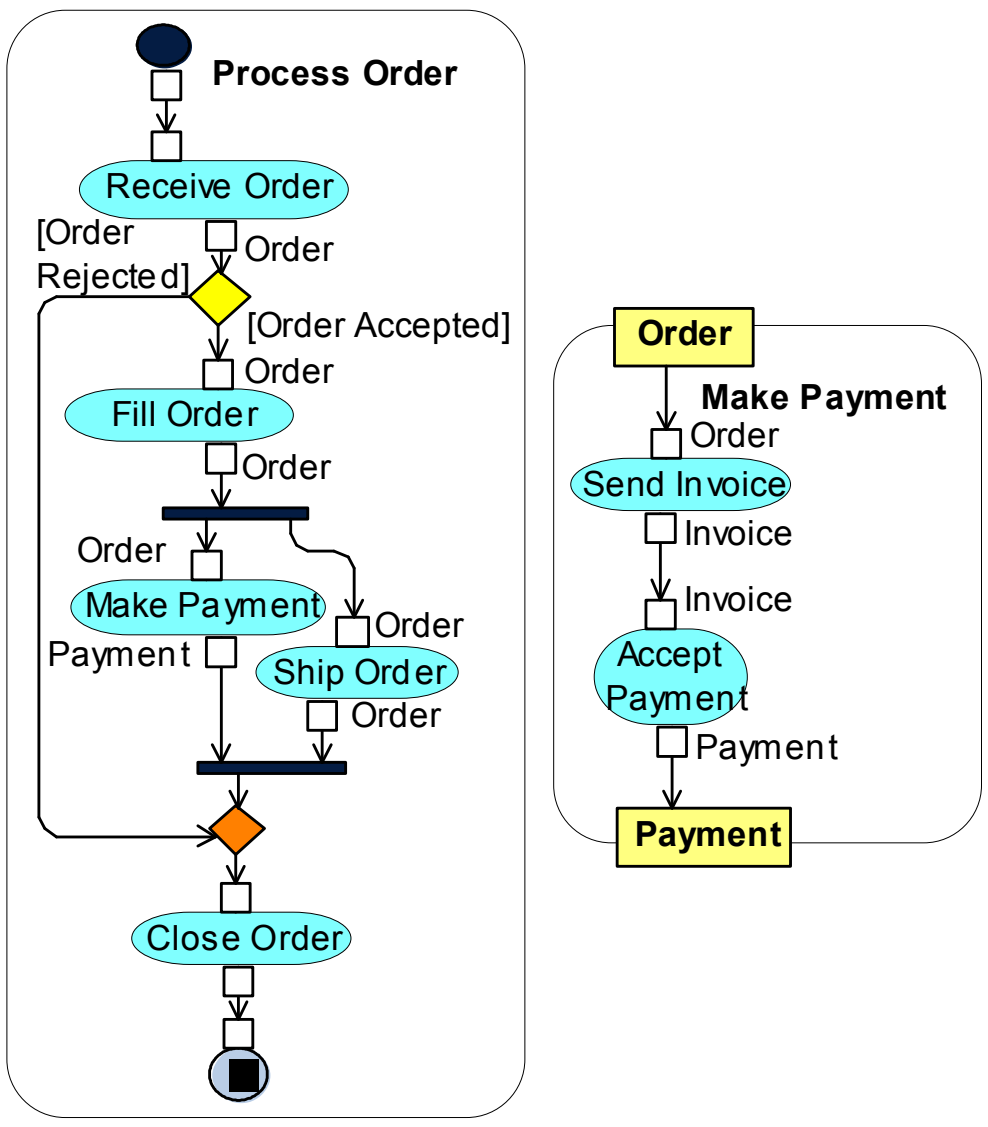

*Fig. 13 Sample activity diagram "Process Order," which invokes "Make Payment"*

Because branching in diagrams is shown explicitly with decision and fork nodes, only one edge can extend from the pin. Guards for edges outgoing from decisions are mutually exclusive and do not change over time. In addition, the following nonstandard connections are disabled:

- The outgoing edge of a ControlNode cannot be an incoming edge for the same ControlNode. This prevents deadlocks between control nodes waiting for input.
- No paths are allowed between *CallBehaviorActions, InitialNodes, FinalNodes* or *ActivityParameterNodes* containing both *ForkNodes* and *JoinNodes*. This is reasonable from the practical point of view, because there is no need to create parallel branches if they are simply joined back without any operation in these branches (i.e., there is no *CallBehaviorAction* between them).

These restrictions in general do not limit the development of natural business processes, but eliminate the "race for tokens" and the undetermined execution of diagrams. Thus, token movement rules are significantly simplified.

## 4.2 General Description of the UML 2.0 Activity Diagram and Proposed Virtual Machine

### 4.2.1 Standard Semantics of Activity Diagrams

In the UML standard, AM semantics is described in a highly distributed manner, where each AD element has its role in AD execution [11,43,44]. Each fork, decision, merge and join node processes the token flow in its own way by "offering" tokens to actions. The "offering" of a token simply means that control nodes make tokens "visible" to actions, and an action is executed "when all of the input pins are offered tokens and accept them all at once, precluding them from being consumed by any other actions" [11]. This means that actions use pull semantics for token processing, and the only active elements in a diagram are the "action engines," which try to fill up their input pins with a fresh sets of tokens to be consumed by these actions.

A "standard ADVM" could be defined with "action engines" as the only active elements and control nodes as "token visibility switches," but it would be highly complicated because visibility rules are obviously non-local (distributed) operations with many more operating and dependent elements.

### 4.2.2 General Principles of the Developed ADVM

The author proposes a different version of the ADVM, in which control nodes ("unstable places," where tokens cannot be located) and edges are "truncated" in paths. As a result, paths connect nodes where tokens can be located ("stable places"), which are usually input and output pins (for actions, initial and final nodes), as well as activity parameter nodes (for activities). Each path has a condition: the guards of its edges are "anded" together. The abovementioned diagram constraints ensure that paths are mutually exclusive. Even if a pin is a start point for several paths, for a particular token only one path is allowed.

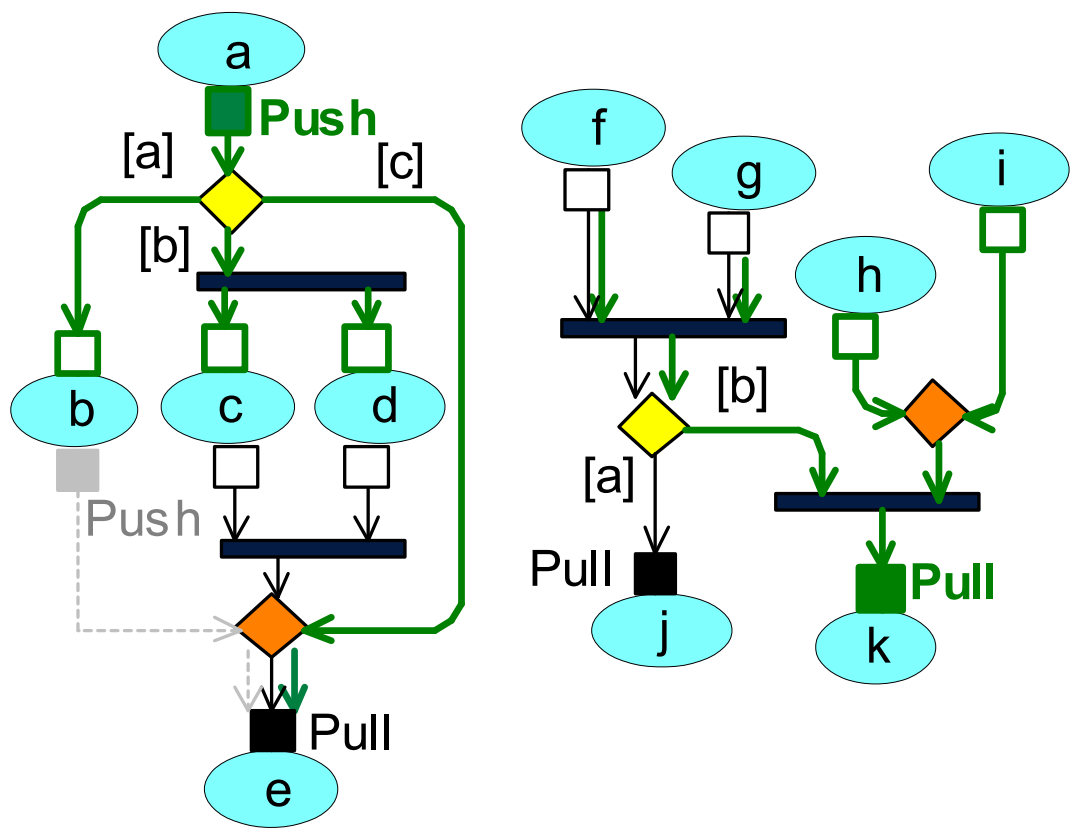

*Fig. 14 Creation of push and pull paths for different connections*

"Stable places" are served by active elements: token engines. Two types of engines are introduced, push and pull engines, as well as push and pull paths. Push paths are those containing only decision, merge and fork nodes, or no control nodes at all. A push path is "serviced" by a push engine in its start node, the corresponding output pin. In the proposed subset, tokens from an output pin can be pushed via push paths independently from each other directly to their destinations, input pins, whenever path conditions permit it. Thus, token movement is very transparent in the push case.

Pull paths are those containing at least one join node, and decisions and merges. Pull paths are serviced by a pull engine at their destination, an input pin. According to AD semantics, the movement of tokens along pull paths having a common destination must be coordinated; only an adequate set of tokens can jointly pass a join node.

The action engine is much simpler than its counterpart in the original standard semantics. Its sole task is to seize one token from each input pin (or an entire group, if this is a pull pin), when a complete set is present and to "consume" this set, and provide output pins.

The main semantic difference between the proposed and the standard action engine is that for the proposed engine, tokens or groups are moved by token engines independently of each input pin, while the standard engine pulls them from output pins "all at once." However, this cannot lead to serious differences in behavior, since real "races for tokens" by several actions are impossible in the proposed subset.

### 4.2.3 Metamodel Extensions and Model Mapping

To formally define an ADVM, a specific AD execution metamodel must be created with all required classes and operations. For each AD, a diagram execution model is created using model transformation.

Fig. 15 shows the original AD and its run-time metamodels (virtual machine) combined; the original classes are light and the new ones are dark. Whenever possible, the corresponding classes in both metamodels are linked by special bidirectional associations (so-called mapping associations, introduced in Chapter 2.3, dashed lines).

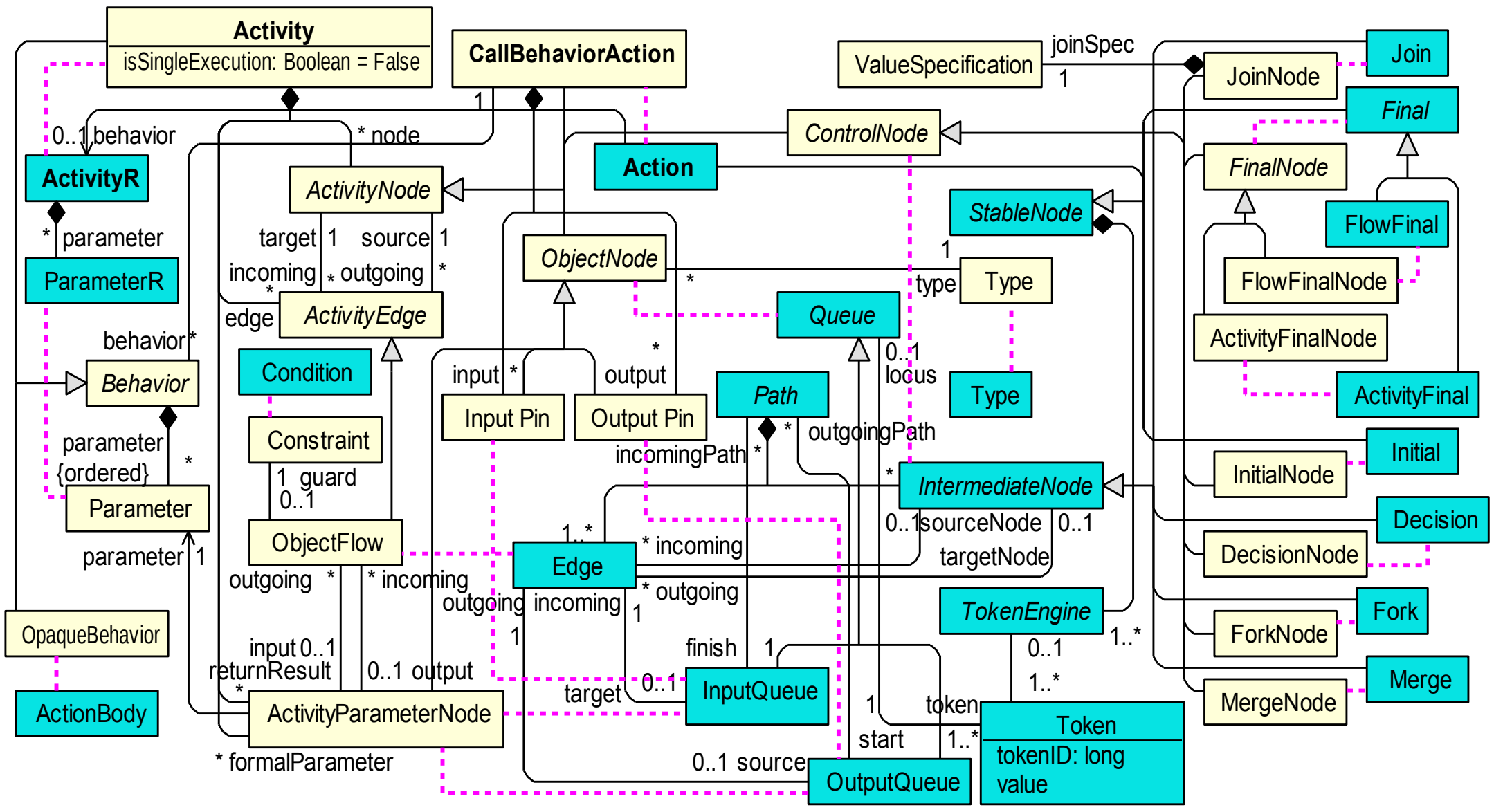

*Fig. 15 Subset of the UML AD and relations to run-time classes*

When an activity is invoked, corresponding run-time class instances are created for the activity instance and all its components. These instances act as the virtual machine executing the given activity. Fig. 15 represents a "general transformation schema," where the mapping associations have a formal semantics in this transformation. In the direction from a definition class to run-time class it means that in the transformation process for each instance of the definition class, one instance of the run-time class should be created. In the opposite direction it shows from which definition instance the run-time instance is created. This information is used when new instances must be created with specific properties that can be retrieved only from the source (definition) model.

Fig. 16 shows the metamodel of the proposed ADVM. This diagram is another view of the metamodel shown in Fig. 15, but with more detailed run-time classes. It shows the complete set of classes, associations and operations required for execution of the proposed VM.

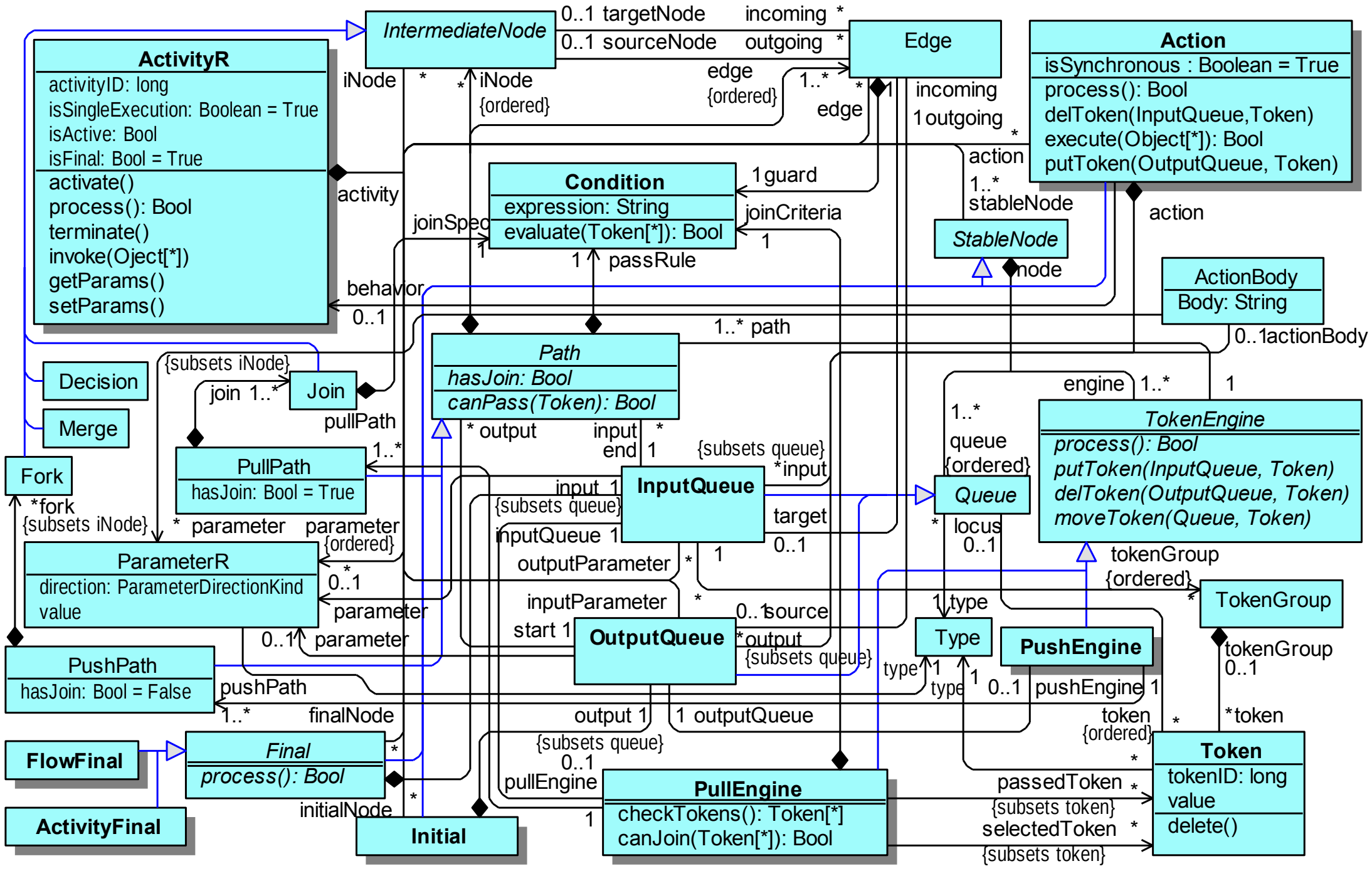

*Fig. 16 Metamodel of the ADVM*

The creation (transformation) of each execution element and its behavior is completely described in the publication. Java pseudocode is used where a procedural approach is better, but OCL constraints are used when a declarative approach is appropriate. As shown in Chapter 5, such a transformation also can be defined with a specialized model transformation language, e.g., MOLA.

Because the proposed VM works differently than the original one on the token movement level, it must be proven that both work the same on an action execution level. Therefore, this research proves that, in the proposed VM, even if movement times for particular tokens differ, the token paths and action execution start and end times fired by tokens in the selected subset are the same.

This proof is based on the following:

- The fact that, in the selected subset of AD elements, both original and proposed VMs work in a determined way.
- Even if, in the proposed VM, a token can arrive at its destination "stable place" earlier than in the original machine, the last token always arrives at its destination at the same time, when the original machine takes all its tokens "at once."

It can be concluded that, on the level of action execution, the proposed VM works the same as the original one, but the proposed machine is much simpler. (In the publication, an entire chapter is devoted to proving this.)

Pseudocode and OCL constraints also demonstrate machine behavior (token movement) graphically and precisely. A machine built in such way is actually the design of a validation or simulation engine. The proposed approach can be used as a basis for developing a workflow management system engine.

# 5 Use of UML and Model Transformations for Workflow Process Definitions

Model driven architecture (MDA) using model driven development (MDD) is taking on an increasingly significant role in the development of business management systems. In turn, model transformations substantially facilitate the development process in MDD style for different system development steps of the system lifecycle.

When using a model-based approach to develop business systems, an exact business modeling language is necessary. As there is no single "killer notation" for workflow modeling, quite often, quick transformation from one modeling language to another is necessary.

The goal of the research was to prove it is possible to define exact and automatically executable model transformations, which allow changing business models from one notation to another without losing model semantics.

The approach is illustrated using two of the most popular modeling languages: UML AD [11] and BPMN [17] (both are currently supported by OMG). Transformations are executed using the MOLA model transformation language. However, the approach is not limited to this specific choice of target notation; a reverse transformation or a completely different transformation using other modeling languages could be treated in a similar way.

The research briefly discusses the workflow aspects that are required in practice, and on this basis, a natural AD profile and appropriate subset of BPMN are proposed. The selection of the proposed AD profile and the subset of BPMN are based on findings performed in previous research (Chapters 2, 3 and 4), with added functionality for business-to-business (B2B) features. The semantics of both languages is analyzed by focusing on distributed business processes, process performers and model execution.

As mentioned in Chapter 2.3, at the moment, the OMG Domain Task Force is working on merging the AD and BPMN languages [72] by relating concepts in these languages using a mapping approach [57]. However, considering the complicated nature of concept relation in these languages, the author shows how this relation can be shown in a more exact and effective way using model transformations.

## 5.1 Languages for Workflow Design and their Role

Obviously, workflow definition requires an easy readable graphical language with clear execution semantics [66,67]. Two modeling languages, UML AD and BPMN, were selected from several reviewed modeling languages to demonstrate the approach. UML AD and BPMN were selected because they best satisfy all workflow definition requirements. Further, execution semantics of the chosen languages is analyzed in this paper, with emphasis on business process specifics and how language graphical aids can represent them.

Because UML ADs are clarified in such way that their elements are more appropriate for embedded systems, some AD tailoring is necessary before it can be used for business process modeling. In other words, a DSL (domain specific language) as a profile of the UML is necessary, where introduced stereotypes hide the found AD deficiencies for practical workflow definition. As the transformation target, the BPMN language is chosen. It is partially supported by tools [65]. But this language has its own set of deficiencies, especially the informal semantics and lack of adequate features for data definition. The usability of BPMN for workflow definition is briefly analyzed in order to identify a relevant subset.

To determine clear execution semantics of both languages in business processes, they are mapped to the sole practically executable process modeling language at the present time: BPEL [18]. In cases where the means of expression defined in the standard are not satisfactory, BPEL is extended using *de facto* industry standard means used by vendors implementing this language.

## 5.2 Adjusting UML Activity Diagrams for Workflow Definition

Several AD profiles for defining workflows have been proposed [52,53,54], but none covers all the required business modeling features: cooperation of distributed processes with messages, data processing, the description of manual task performers and the ability to execute a process model. Therefore, the proposed AD is specially tailored. A list of recommended AD features for workflow definition is selected, retaining the original semantics of AD elements, and including some stereotypes that add missing properties and constraints required for distributed workflows.

Fig. 17 shows a workflow example with two activities illustrating the chosen elements. (This and all further diagrams are modeled by using the GMF (EBM) tool [89], where additional editors for UML AD profile and BPMN languages were developed.) These diagrams show practically all elements chosen for workflow modeling. Most of the elements are chosen from previous research (Chapters 2, 3 and 4). Elements typical for B2B cooperation are also added.

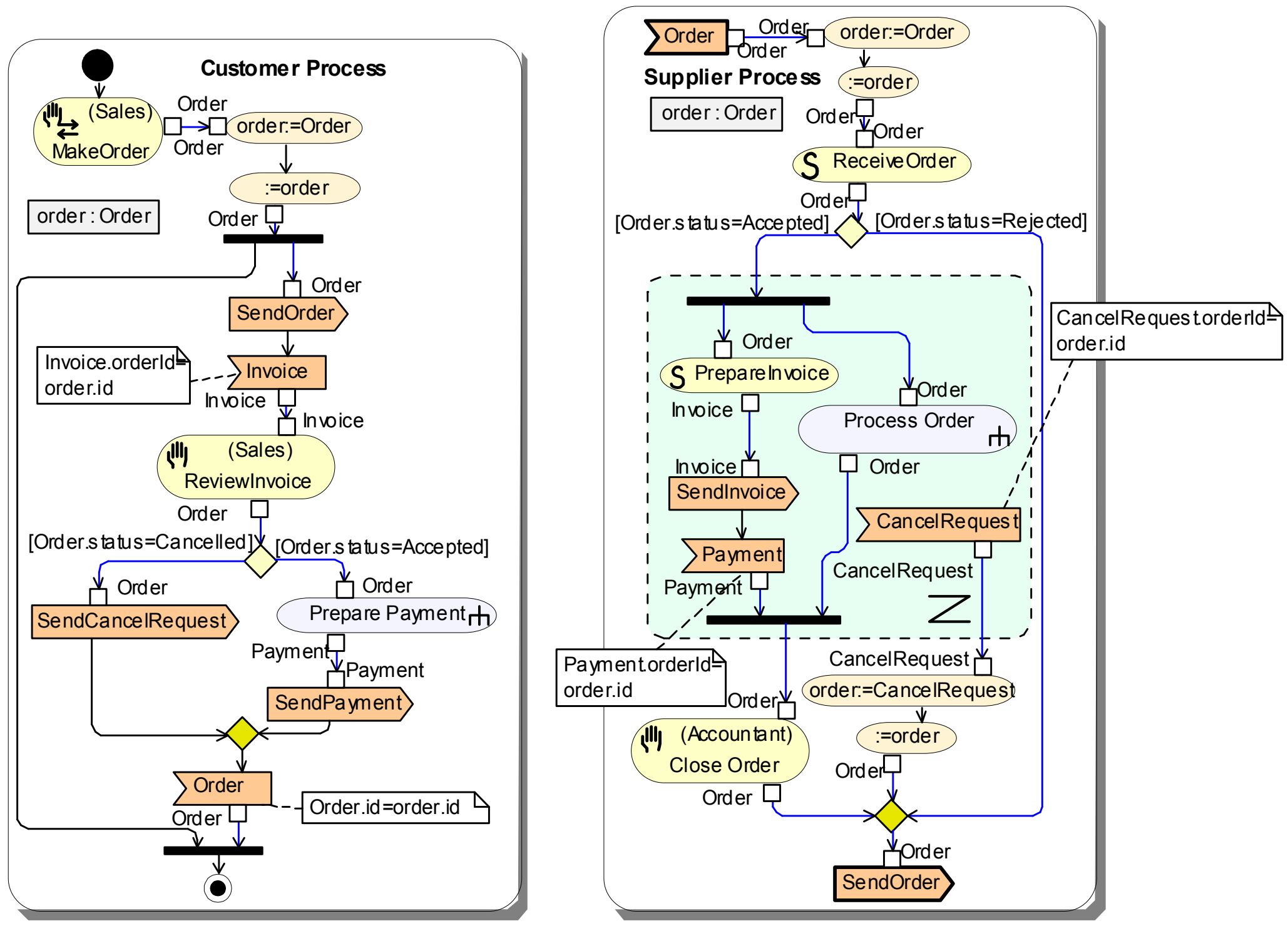

*Fig. 17 Example of a process using UML ADs*

New elements are *SendSignal* and *AcceptEvent* actions (convex and concave flags, respectively), and stereotypes for different kinds of actions. The element `order : Order` is a variable definition of type `Order`, with the scope of activity `Supplier Process`. Write variable actions are presented as normal assignments to the variable, and simple OCL syntax is used for expressions.

*AcceptEvent* actions with data based guard conditions for the outgoing object flow are specific patterns for locating the relevant process instance at message reception, which is equivalent to the use of explicit correlation sets in BPEL.

Using stereotypes, the semantics of some AD elements is adjusted to meet the requirements of distributed workflow modeling. The metamodel in Fig. 18 illustrates the introduced AD stereotypes:

- ❑ *MainProcess* (stereotype for Activity) is a separate workflow process (executed by an individual workflow engine). Graphically, it is shown as a shadowed activity.
- ❑ *Performer* (stereotype for Partition) represents a performer of a manual or user action. Its `represents` association must reference a class with the *Position* or *OrgUnit* stereotype. It is shown as a compartment of the action.
- ❑ *WebService* (stereotype for Component) is used to describe web service attributes. *CallServiceTask* actions, which invoke operations within this service, get their technical parameters from *WebService.*
- ❑ *IntermSSAction* means sending a specified signal to a web service (it is shown as a simple convex flag). *EndSSAction* (shown as a convex flag with a bold border) means sending a signal as the final action of the activity.
- ❑ *ForEach* is a stereotype for *LoopNode.* It has an additional `ForEach.collection` association that references the `ValuePin`. It is introduced for iterators (the requirement for iterators in business modeling languages also mentioned in [66]). It is the only place where heavyweight extension is used, to preserve metamodel compatibility.

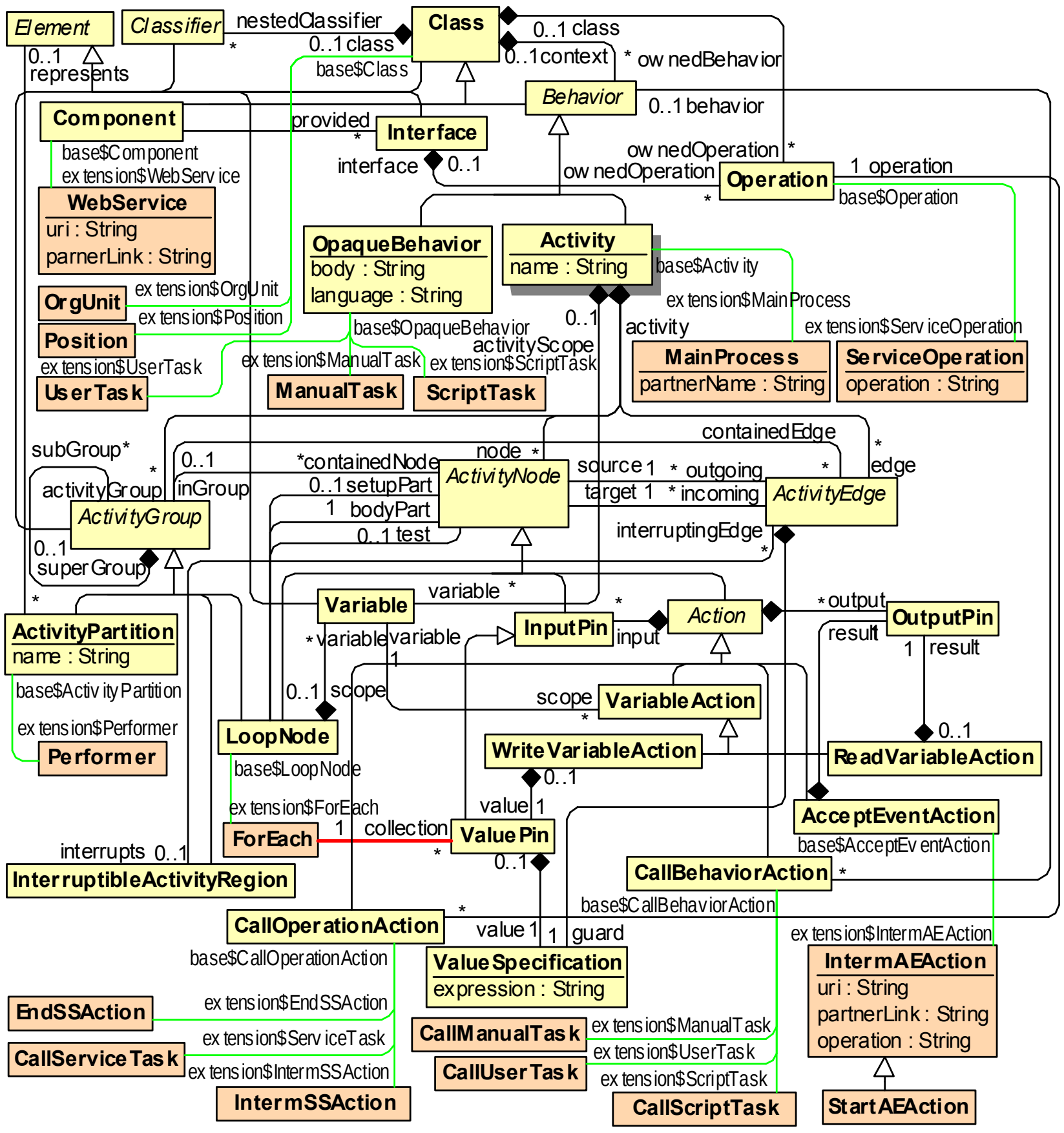

*Fig. 18 Fragment of the AD metamodel (source of model transformation)*

Fig. 18 shows a UML AD metamodel fragment that is "flattened" because redundant abstract superclasses are eliminated, and it has "applied" the proposed workflow modeling profile. According to the MOF standard, profiled classes are linked to main classes with specific associations (e.g., `extension$Position <-> base$Class`). Instances of this metamodel are the source for transformations described in Section 4.4.

## 5.3 BPMN Diagrams as Another Notation

As noted at the start of Chapter 5, BPMN is also a widely used language for workflow definition. As BPMN also has redundancy and some deficiencies, a BPMN subset is proposed in this paper. All kinds of *Gateways* (diamonds), all types of *Tasks* and *Subprocesses* (rounded rectangles), *Start* and *End* events, and *IntermediateEvents* (circles) attached to the boundary of an *Activity* ("interrupt construct") are chosen because they all have a natural semantics and mapping to BPEL.

Fig. 19 shows a process example in the chosen subset of a BPMN notation. Both processes shown here are precise analogues of processes shown in Fig. 17, which are obtained by applying practical MOLA language transformations in the GMF tool, and using the tool's automatic layout possibilities.

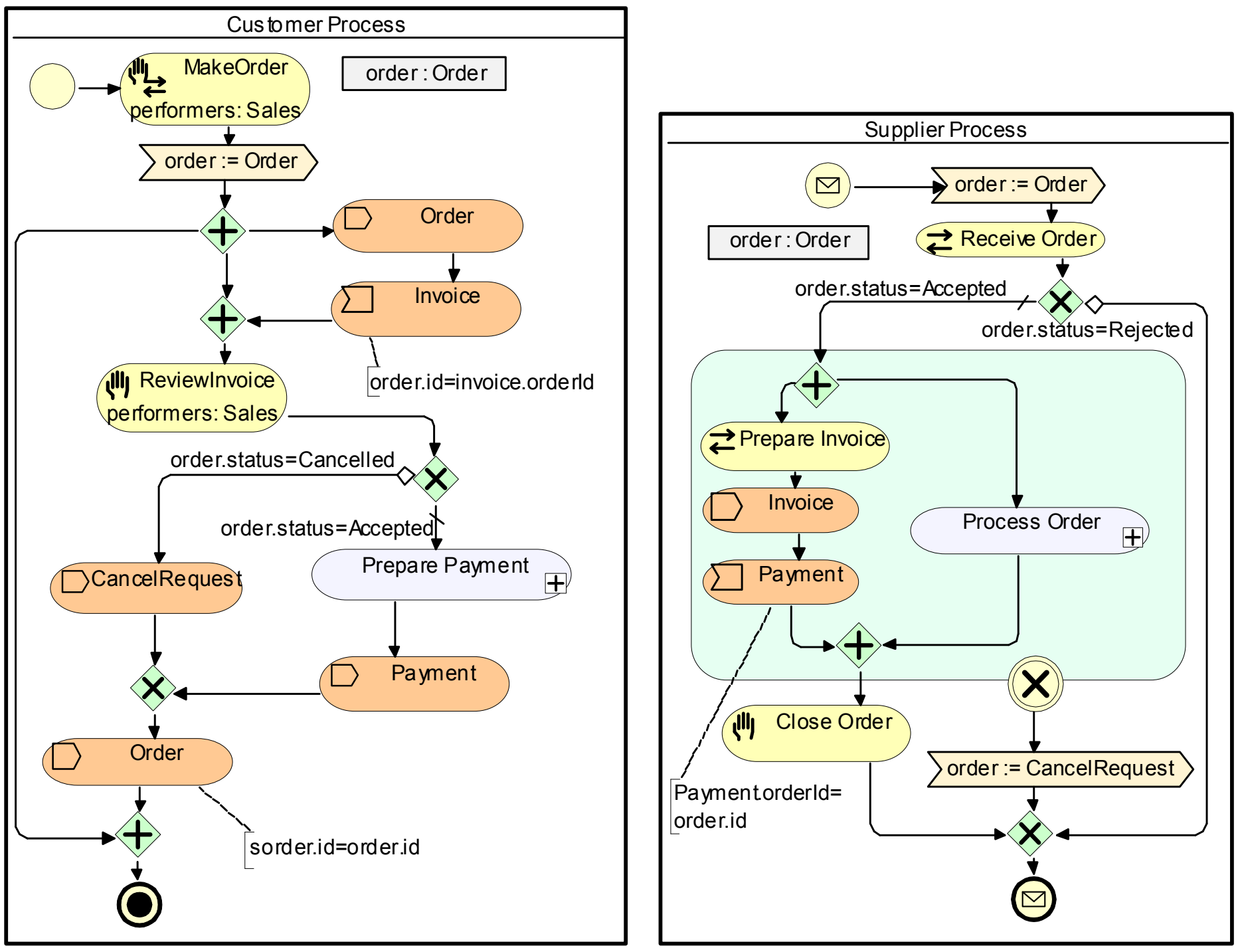

*Fig. 19 Process example in BPMN*

For receiving messages in non-interrupt situations, *ReceiveTasks* are preferred. For sending messages in non-interrupt situations, *SendTasks* are preferred. In both cases, the tasks are represented by rounded rectangles with concave and convex flags inside.

While the implicit BPMN metamodel is quite acceptable, the BPMN graphical notation lacks some important elements. Therefore, a notation for following elements is introduced:

- A stereotyped (with icons) notation for task types and an explicit compartment for task performers are used.
- Properties are represented as rectangles containing `name:type`.
- To make the data aspect visible, assignments are represented as large arrows containing textual assignment statements.

Similarly to UML ADs, Fig. 20 shows a fragment of the proposed BPMN subset. The instances of these elements appear as the transformation result.

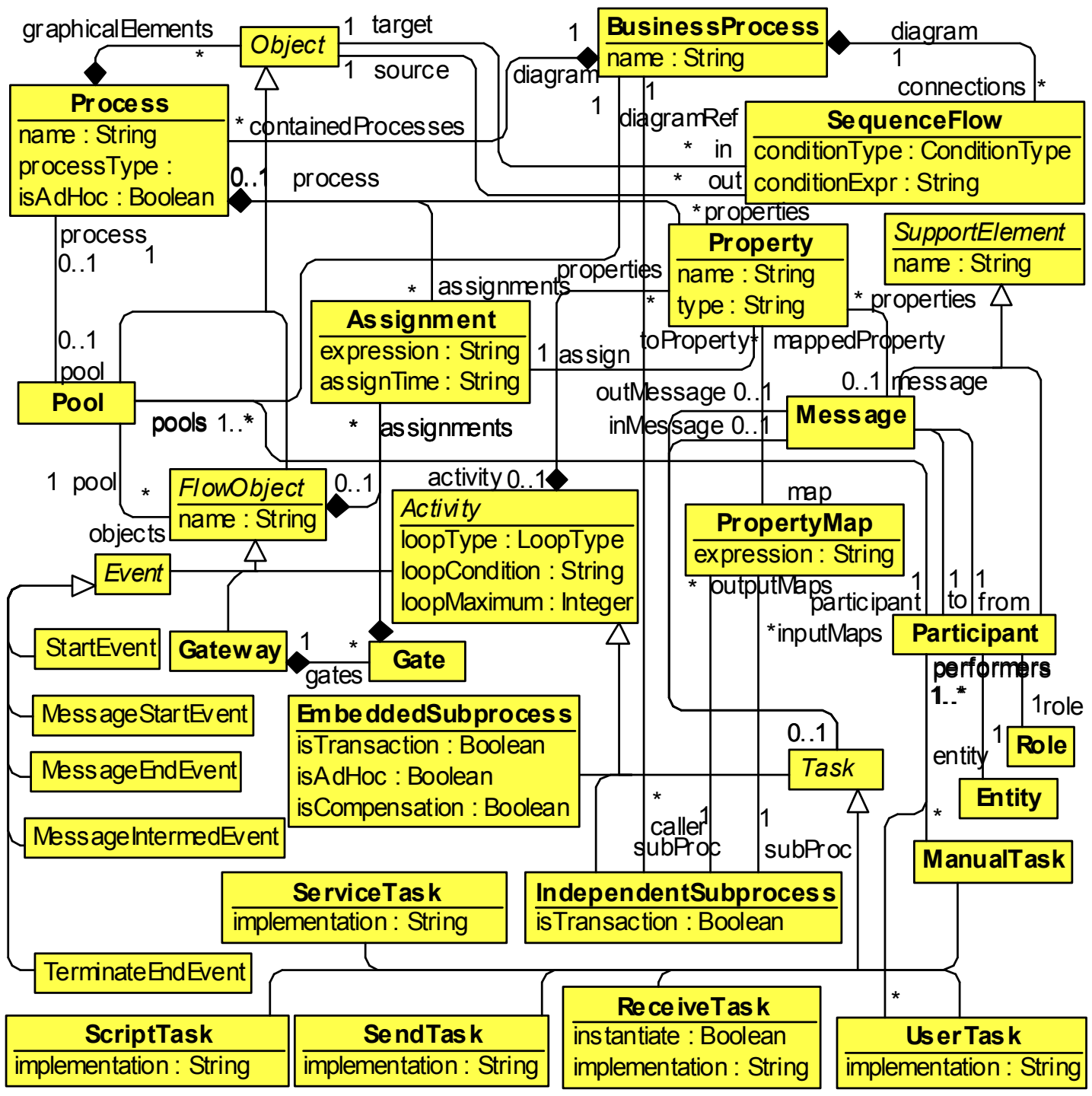

*Fig. 20 Fragment of the BPMN metamodel (target for transformation)*

## 5.4 AD to BPMN Transformation

Although there are several research papers that mention formal process transformations [62,67], no existing formalized transformations from UML ADs to BPMN supporting all workflow related aspects were found. Therefore, the author uses existing AD-to-BPMN mappings [68,69,70,71,76,77] and extends them for data flows, variables, assignments and task performers.

To refine the mapping, in this paper an illustrative fragment of the formal transformation generating a BPMN model from an AD model is provided. The transformation is written in the MOLA language [82]. Fig. 18 and Fig. 20 represent the source and target metamodels of the transformation, respectively, but Fig. 21 illustrates the main program of the transformation, which transforms each *Activity* to a new BPMN Process and Pool, and invokes subprograms transforming all other elements for this *Activity*.

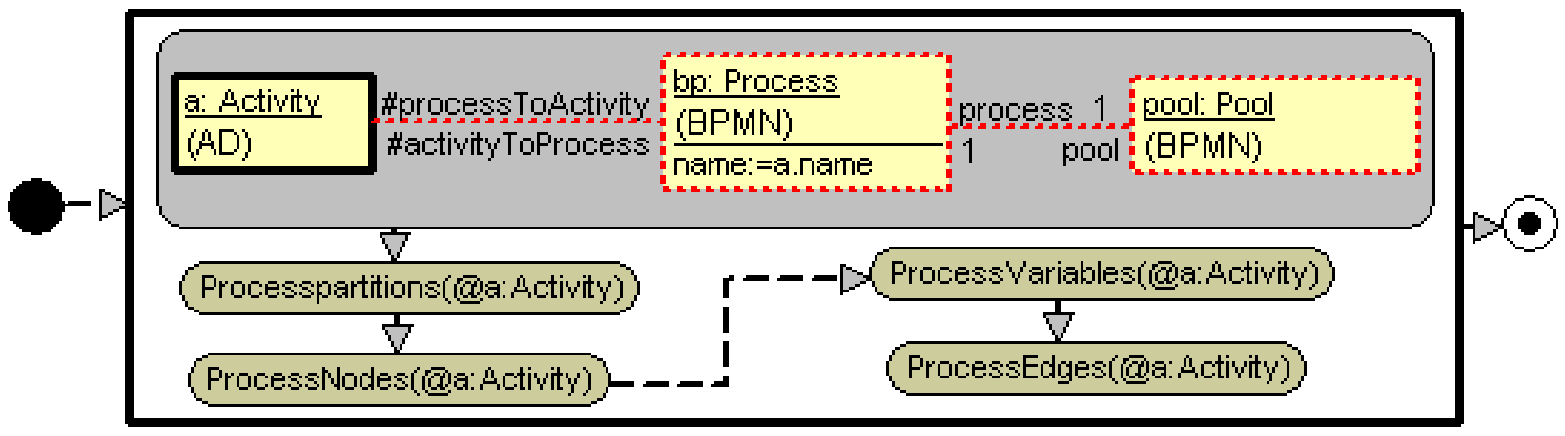

*Fig. 21 Main transformation program of AD to BPMN*

The MOLA program formally and completely describes the concept mapping between these two notations. As shown in Fig. 21, several classes and associations with context are required, which is quite complicated, and cannot be clearly described simply by using mapping associations. Model transformation is the best method for defining a complicated mapping between two notations, while preserving the semantics.

Transformation is performed using the GMF (EBM) tool, using the MOLA language editor and with developed editors for the AD and BMPN languages. The implementation shows the efficiency of the approach: small models with several tens of classes were transformed in a few seconds on a standard personal computer. On the basis of this research, a new model transformation framework is under development. It uses an in-memory repository for model transformations, and this allows executing model transformations for industrial needs.

# 6 Conclusion

The goal of the thesis is to analyze and use a metamodeling approach to define exact (i.e., executable) business processes. Some of the business process metamodeling ideas developed by the author are already implemented in existing modeling tools, but others can be used for the development of new tools. The following is a list of the author’s ideas concerning the latest business process modeling events:

- The "notation independent" business metamodel proposed at the start of the research shows business concepts and their relations, and it served as the basis for all further investigations. The proposed ideas have some similarity to the OMG Business motivation model [51] and Process definition model [50] standard drafts, which were proposed later. However, in the author's metamodel, the part used for process inputs, outputs and task performers is more detailed than in the standard drafts. On the basis of the slightly modified metamodel, the author has developed editors for the UML AD and BPMN languages in the GMF (EBM) tool [89].
- The author's proposal for concept mapping from one domain to several presentations is used in the GMF tool, where the same business model can be shown in several similar modeling languages. The idea of representing similar languages as different views of a "canonical form" has recently been reflected in the OMG initiative, which attempts to join the UML AD and BPMN modeling languages in a single domain [57].
- The transformation examples of the MOLA [82] language executed during the research represent a considerable contribution toward the validation, testing and demonstration of the MOLA tool. The idea of business model transformations from one language to other is realized in the MOLA tool, allowing several users to use the most convenient modeling language in different stages of developing business management systems.
- At this time, the University of Latvia IMCS has started developing a new generation modeling tool platform based on model transformations (TTF [90]), where model transformations are realized in the MOLA language. Development of specialized modeling tools for domain-specific languages (DSL) will be one of the possible usages of this new generation tool platform. For example, it will be possible to build different business modeling tools on the platform. The author's developed UML AD profile and subset of the BPMN language can be used as a basis for developing process modeling editors on this platform. In turn, the author's developed ADVM and business measure framework can be used to develop a process simulation engine. In this way, all necessary functionality for business process modeling and simulation will be provided.

# 7 References

## 7.1 Referenced Papers by the Author

## 7.2 Other Publications by the Author

## 7.3 Other Sources